\documentclass[%
 aip,
 amsmath,amssymb,
reprint,%
floatfix,
]{revtex4-1}

\usepackage{graphicx}
\usepackage{braket}
\usepackage{csquotes}
\usepackage{algpseudocode}

\newcommand{\indi}[0]{I}
\newcommand{\indj}[0]{J}
\newcommand{\indp}[0]{p}
\newcommand{\indq}[0]{q}

\newcommand{\then}[1]{N_\mathrm{#1}}
\newcommand{\thei}[0]{\mathrm{i}}

\newcommand{\ther}[1]{\mathbf{R}_{#1}}
\newcommand{\therdot}[1]{\dot{\mathbf{R}}_{#1}}
\newcommand{\therdotsq}[1]{\dot{R}^2_{#1}}
\newcommand{\therddot}[1]{\ddot{\mathbf{R}}_{#1}}
\newcommand{\thep}[1]{\mathbf{P}_{#1}}
\newcommand{\thepdot}[1]{\dot{\mathbf{P}}_{#1}}
\newcommand{\thek}[1]{\mathbf{K}_{#1}}
\newcommand{\thekdot}[1]{\dot{\mathbf{K}}_{#1}}
\newcommand{\thepi}[1]{\boldsymbol{\Pi}_{#1}}
\newcommand{\thepidot}[1]{\dot{\boldsymbol{\Pi}}_{#1}}
\newcommand{\thechi}[2]{\boldsymbol{\chi}_{#1} (#2)}
\newcommand{\thechidot}[2]{\dot{\boldsymbol{\chi}}_{#1} (#2)}
\newcommand{\theom}[3]{\boldsymbol{\Omega}_{#1#2} (#3)}
\newcommand{\thew}[2]{\mathbf{W}_{#1} (#2)}
\newcommand{\thewdot}[2]{\dot{\mathbf{W}}_{#1} (#2)}
\newcommand{\them}[1]{\ifthenelse{\equal{#1}{}}{\mathbf{M}}{M_{#1}}}
\newcommand{\thez}[1]{Z_{#1}}
\newcommand{\thea}[2]{\mathbf{A} (#2{}_#1{})}

\newcommand{\theza}[2]{Z_{#1} e \mathbf{A} (#2{}_#1{})}
\newcommand{\thezadot}[2]{Z_{#1} e \dot{\mathbf{A}} (#2{}_#1{})}
\newcommand{\thertwo}[1]{\mathbf{R}'_{#1}}
\newcommand{\thertwodot}[1]{\dot{\mathbf{R}}'_{#1}}
\newcommand{\thertwoddot}[1]{\ddot{\mathbf{R}}'_{#1}}
\newcommand{\theptwo}[1]{\mathbf{P}'_{#1}}

\newcommand{\thepitwo}[1]{\boldsymbol{\Pi}'_{#1}}
\newcommand{\theb}[1]{\ifthenelse{\equal{#1}{}}{\mathbf{B}}{B_{#1}}}
\newcommand{\thef}[3]{\mathbf{F}_{#1}^\mathrm{#2} #3}
\newcommand{\theop}[3]{#1_\mathrm{#2}^\mathrm{#3}}
\newcommand{\theopbar}[3]{\bar{#1}_\mathrm{#2}^\mathrm{#3}}
\newcommand{\thelag}[0]{\mathcal{L}}
\newcommand{\thelagbar}[0]{\bar{\mathcal{L}}}
\newcommand{\thewf}[3]{
    \ifthenelse{\equal{#1}{nuc}}
        {\ifthenelse{\equal{#3}{1}}{\psi_{#2} (\ther{},t)}{\psi_{#2}}}
        {\ifthenelse{\equal{#3}{1}}{\phi_{#2} (\mathbf{r};\ther{})}{\phi_{#2}}}
}
\newcommand{\thewftot}[1]{\ifthenelse{\equal{#1}{1}}{\Psi (\ther{},\mathbf{r},t)}{\Psi}}
\newcommand{\thee}[2]{#1_\mathrm{#2}}
\newcommand{\thewvn}[1]{\tilde{\nu}_{#1}}
\newcommand{\thephi}[2]{\Phi_\mathrm{#1} #2}
\newcommand{\theomega}[1]{\boldsymbol{\omega}_\mathrm{#1}}
\newcommand{\theg}[1]{\mathbf{G}}

\begin{document}

\title[]{Ab-Initio Molecular Dynamics with Screened Lorentz Forces. Part II.\\ Efficient Propagators and Rovibrational Spectra in Strong Magnetic Fields}

\author{Laurens D. M. Peters}
\email{laurens.peters@kjemi.uio.no}
\affiliation
{Hylleraas Centre for Quantum Molecular Sciences,  Department of Chemistry, 
University of Oslo, P.O. Box 1033 Blindern, N-0315 Oslo, Norway}
\author{Tanner Culpitt}
\affiliation
{Hylleraas Centre for Quantum Molecular Sciences,  Department of Chemistry, 
University of Oslo, P.O. Box 1033 Blindern, N-0315 Oslo, Norway}
\author{Laurenz Monzel}
\affiliation
{Hylleraas Centre for Quantum Molecular Sciences,  Department of Chemistry, 
University of Oslo, P.O. Box 1033 Blindern, N-0315 Oslo, Norway}
\affiliation{Karlsruhe Institute of Technology (KIT), Institute of Physical Chemistry, KIT Campus South, P.O. Box 6980, D-76049 Karlsruhe, Germany}
\author{Erik I. Tellgren}
\affiliation
{Hylleraas Centre for Quantum Molecular Sciences,  Department of Chemistry, 
University of Oslo, P.O. Box 1033 Blindern, N-0315 Oslo, Norway}
\author{Trygve Helgaker}
\affiliation
{Hylleraas Centre for Quantum Molecular Sciences,  Department of Chemistry, 
University of Oslo, P.O. Box 1033 Blindern, N-0315 Oslo, Norway}

\date{\today}

\begin{abstract}
Strong magnetic fields have a large impact on the dynamics of molecules. In addition to the changes of the electronic structure, the nuclei are exposed to the Lorentz force with the magnetic field being screened by the electrons. In this work, we explore these effects using \textit{ab-initio} molecular dynamics simulations based on an effective Hamiltonian calculated at the Hartree--Fock level of theory. To correctly include these non-conservative forces in the dynamics, we have designed a series of novel propagators that show both good efficiency and stability in test cases. As a first application, we analyze simulations of He and H$_2$ at two field strengths characteristic of magnetic white dwarfs (0.1\,$\theb{0} = 2.35 \times 10^4$\,T and $\theb{0} = 2.35 \times 10^5$\,T). While the He simulations clearly demonstrate the importance of electron screening of the Lorentz force in the dynamics, the extracted rovibrational spectra of H$_2$ reveal a number of fascinating features not observed in the field-free case: couplings of rotations/vibrations with the cyclotron rotation, overtones with unusual selection rules, and hindered rotations that transmute into librations with increasing field strength. We conclude that our presented framework is a powerful tool to investigate molecules in these extreme environments.
\end{abstract}

\maketitle

\section{Introduction}

The magnetic field strength is a physical quantity that varies extremely throughout the universe. While the Earth's magnetic field ranges from 25 to 65\,$\mu$T,\cite{Finlay2010}magnetic white dwarfs can develop fields up to $10^5$\,T and neutron stars even up to $10^{11}$\,T.\cite{Ferrario2015,Reisenegger2003} At such field strengths, the physics and chemistry of atoms and molecules are heavily affected by the magnetic field.\cite{Garstang1977,Lai2001} Of particular interest is the region on the order of one atomic unit field strength $\theb{0} =2.35 \times 10^5$\,T, where the interaction energies of the electrons with the magnetic field are of the same order of magnitude as the Coulomb interactions that dominate chemistry at zero field strength. 

As field strengths on the order of $B_0$ are so far not accessible via experiments,\cite{Nakamura2018} theoretical calculations remain the only tools to investigate atoms\cite{Kravchenko1996,Ivanov1998,Ivanov1999,Jones1999,Ivanov2000,Al-Hujaj2000,Becken2002,Al-Hujaj2004,Thirumalai2009,Thirumalai2014} and molecules\cite{Schmelcher1990,Ozaki1993,Kappes1994,Detmer1997,Detmer1998} under these conditions. It has been shown that London atomic orbitals,\cite{London1937,Hameka1958,Ditchfield1976,Helgaker1991} which contain a field-dependent phase factor, are very useful for this purpose in that they ensure exact gauge-origin invariance in the calculations. The derived one- and two-electron integrals\cite{Tellgren2008,Tellgren2012,Reynolds2015,Irons2017,Williams-Young2020,Pausch2020} give access to energies and properties of various systems in strong magnetic field at the Hartree--Fock (HF) level of theory.\cite{Tellgren2008,Tellgren2009,Tellgren2012} This framework was then expanded to  full-configuration-interaction,\cite{Lange2012,Austad2020} coupled-cluster,\cite{Stopkowicz2015} and linear-response\cite{Sen2019} calculations to get deeper and more accurate insights into these systems---revealing, for example, a novel paramagnetic bonding mechanism\cite{Lange2012}.

Despite these developments in the field of electronic-structure theory in strong magnetic fields, only very little is known about molecular dynamics in these environments. The main challenge is that not only is the electronic structure affected, leading to a different potential energy surface than in the field-free case, but the atoms are subject an additional force, the Lorentz force, which influences their motion. To account for the Lorentz force, Spreiter and Walter\cite{Spreiter1999} modified the well-known velocity Verlet\cite{Verlet1967,Swope1982} propagator. The resulting algorithm based on a Taylor expansion has been implemented in prominent classical molecular dynamics program packages\cite{DellaValle2017,Khajeh2020} and has been used in various applications~\cite{Al-Haik2006,Chang2006,Daneshvar2020}.

Nonetheless, these simulations were conducted using classical force fields, so the effect of the magnetic field on the electronic structure as well as the electronic screening of the magnetic field acting on each nucleus was neglected. While the theoretical foundation for \textit{ab-initio} molecular dynamics in a strong magnetic field including this screening was laid already thirty years ago by Schmelcher and coworkers\cite{Schmelcher1988,Schmelcher1989,Schmelcher1997}, there is (to our knowledge) only one publication that reports on such simulations. Carrying out Hartree--Fock calculations in a minimal Slater basis, Ceresoli, Marchetti, and Tosatti demonstrated the importance of electronic screening of the magnetic field acting on the nuclei by simulating the motion of H$_2$ perpendicular to the magnetic field vector.~\cite{Ceresoli2007}

This is the second of two works aiming at performing and investigating accurate dynamics of molecules in a strong magnetic field. Here, we use the electronic-structure program package {\sc London}\cite{London} to conduct \textit{ab-initio} molecular dynamics simulations with proper account of screening of the strong magnetic field by the electrons (denoted Berry screening from now on) with the Berry curvature discussed in Part I.~\cite{Culpitt2021} For this purpose, we develop a series of new propagators for the efficient integration of the nuclear equations of motion under the influence of an electronically screened Lorentz force, based on the work of Tao on the propagation of Hamilton's equations of motion with a nonseparable Hamiltonian\cite{Tao2016}. Such propagators are then used to simulate the motion of He and H$_2$ at two different field strengths ($0.1\,\theb{0}$ and $1.0\,\theb{0}$) at the Hartree--Fock level of theory. By extracting rovibrational spectra from the sampled momenta, we investigate for the first time the  effect of a strong magnetic field on molecular vibrations and rotations and, in particular, the role of the Berry screening.  

We begin with presenting the theory in Section~II, deriving the effective Hamiltonian in a magnetic field, the equations of motion, and their integration. In Section~III, we summarize the computational details. The validation and performance tests of the propagators are found in Section~IV, while Section~V presents and analyzes the dynamics and spectra of He and H$_2$. Conclusions and an outlook are given in Section~VI.

\section{Theory}

Throughout this work, $\indi$ and $\indj$ will serve as indices for the $\then{nuc}$ nuclei. We use the notation $\them{\indi}$, $\thez{\indi}$, $\ther{\indi}$, and $\thep{\indi} = - \thei \hbar \partial/\partial \ther{\indi}$ for the  mass, charge, position operator, and canonical momentum operator of nucleus $\indi$, whereas $\thea{\indi}{\ther{}} = \frac{1}{2} \theb{} \times (\ther{\indi} - \theg{})$ is the external vector potential of the uniform magnetic field $\theb{}$ and origin $\theg{}$. The vectors of collective electronic and nuclear coordinates are denoted by $\mathbf{r}$ and $\mathbf{R}$, respectively. 

\subsection{The Effective Hamiltonian in a Magnetic Field}

To derive the effective Hamiltonian in a uniform magnetic field, we start from the general Hamiltonian
\begin{equation}
\theop{H}{}{} = \theop{T}{\mathrm{nuc}}{} + \theop{H}{\mathrm{el}}{} ,
\label{ham_000}
\end{equation}
where $\theop{H}{\mathrm{el}}{}$ contains all electronic contributions including the nuclear repulsion, whereas $\theop{T}{\mathrm{nuc}}{}$ is the nuclear kinetic energy operator
\begin{equation}
\theop{T}{\mathrm{nuc}}{} = 
\sum \limits_{\indi=1}^{\then{nuc}} \dfrac{1}{2\them{\indi}} [\thep{\indi} - \theza{\indi}{\ther{}}]^2 .
\label{ham_001}
\end{equation}
Expanding the total wavefunction ($\Psi$) in an infinite series of time-independent eigenstates $\thewf{el}{\indp}{0}$ of the electronic Hamiltonian $\theop{H}{\mathrm{el}}{}$  with time-dependent expansion coefficients $\thewf{nuc}{\indp}{0}$,
\begin{equation}
\thewftot{1} = 
\sum \limits_{\indp} \thewf{nuc}{\indp}{1} \thewf{el}{\indp}{1} ,
\label{ham_002}
\end{equation}
we can set up the Schr\"odinger equation:
\begin{align}
\theop{H}{}{}  \ket{\thewftot{1}} &= 
\sum \limits_{\indp} [\theop{T}{nuc}{} + \theop{H}{el}{}] \ket{\thewf{nuc}{\indp}{1} \thewf{el}{\indp}{1}}
\nonumber \\
 &= \thei \hbar \sum \limits_{\indp} \dfrac{\partial}{\partial t} \ket{\thewf{nuc}{\indp}{1} \thewf{el}{\indp}{1}} 
 \label{ham_003}
\end{align}
From now on, we suppress the arguments of the wavefunctions to ease the reading of the equations. Multiplying with the electronic state $\bra{\thewf{el}{\indq}{0}}$ from the left,
\begin{equation}
\sum \limits_{\indp}
\bra{\thewf{el}{\indq}{0}} \theop{T}{nuc}{} + \theop{H}{el}{} \ket{ \thewf{nuc}{\indp}{0} \thewf{el}{\indp}{0}} = 
\thei \hbar \sum \limits_{\indp} \braket{\thewf{el}{\indq}{0} | \dfrac{\partial}{\partial t}  \thewf{nuc}{\indp}{0}  \thewf{el}{\indp}{0}} ,
\label{ham_004}
\end{equation}
and recalling that the electronic eigenstates $\phi_p$ are time-independent and orthonormal, we may simplify the matrix elements in the manner
\begin{align}
\sum \limits_{\indp}
\bra{\thewf{el}{\indq}{0}} \thei \hbar \dfrac{\partial}{\partial t} \ket{ \thewf{nuc}{\indp}{0}  \thewf{el}{\indp}{0}} &= \thei \hbar \dfrac{\partial}{\partial t} \ket{ \thewf{nuc}{\indq}{0}} 
, \label{ham_005} \\
\sum \limits_{\indp}
 \bra{\thewf{el}{\indq}{0}} \theop{H}{el}{} \ket{ \thewf{nuc}{\indp}{0}  \thewf{el}{\indp}{0}} &= \thee{U}{BO} (\ther{}) \ket{ \thewf{nuc}{\indq}{0}} 
, \label{ham_006}
\\
\sum \limits_{\indp}
\bra{\thewf{el}{\indq}{0}} \theop{T}{nuc}{} \ket{ \thewf{el}{\indp}{0}  \thewf{nuc}{\indp}{0}} &= 
\bra{\thewf{el}{\indq}{0}} \theop{T}{nuc}{} \ket{\thewf{el}{\indq}{0}  \thewf{nuc}{\indq}{0}} + \theop{T}{nac}{}
\nonumber \\ &= 
[\theop{T}{eff}{} + \thee{U}{d} (\ther{})]\ket{ \thewf{nuc}{\indq}{0}} + \theop{T}{nac}{} ,
\label{ham_007}
\end{align}
where the nondiagonal, nonadiabatic term $\theop{T}{nac}{}$ couples the electronic state $\thewf{el}{\indq}{0}$ to other electronic states. Invoking the Born--Huang approximation,\cite{Born1954,Ballhausen1972} this coupling is henceforth neglected by setting $\theop{T}{nac}{}$ to zero.
We omit the index $\indq$ from now on to emphasize that $\thewf{nuc}{}{0}$ and $\thewf{el}{}{0}$ refer to ground-state wavefunctions. In the above equations, we have also introduced the Born--Oppenheimer potential energy (which includes the influence of the magnetic field on the electronic structure)
\begin{equation}
\thee{U}{BO} =  
\bra{\thewf{el}{\indq}{0}} \theop{H}{el}{} \ket{\thewf{el}{\indq}{0}},
\label{ham_008}
\end{equation}
the effective nuclear kinetic energy operator\cite{Ceresoli2007}
\begin{equation}
\theop{T}{eff}{}  = 
\sum \limits_{\indi=1}^{\then{nuc}} \dfrac{1}{2 \them{\indi}} [\thep{\indi} - \theza{\indi}{\ther{}} + \thechi{\indi}{\ther{}} ]^2 ,
\label{ham_009}
\end{equation}
and the diagonal nonadiabatic potential energy correction
\begin{equation}
\thee{U}{d} =
\sum \limits_{\indi=1}^{\then{nuc}} \dfrac{1}{2 \them{\indi}} [\braket{\thep{\indi} \thewf{el}{}{0}| \thep{\indi} \thewf{el}{}{0}} - |\thechi{\indi}{\ther{}}|^2] ,
\label{ham_010}
\end{equation}
where $\braket{\thep{\indi} \thewf{el}{}{0}| \thep{\indi} \thewf{el}{}{0}}$ is the diagonal Born--Oppenheimer correction and $\thechi{\indi}{\ther{}}$ the geometric vector potential\cite{Yin1994,Resta2000}
\begin{equation}
\thechi{\indi}{\ther{}} = \braket{\thewf{el}{}{0} | \thep{\indi} \thewf{el}{}{0}} =  - \thei \hbar \braket{\thewf{el}{}{0}|\dfrac{\partial}{\partial \ther{\indi}} \thewf{el}{}{0}} .
\label{ham_011}
\end{equation}
If we assume that $\thee{U}{d}$ can be neglected in regions where the Born--Oppenheimer approximation is valid, Eq.\,\eqref{ham_004} can be rewritten as
\begin{equation}
\theop{H}{eff}{} \ket{\thewf{nuc}{}{0}} = 
[\theop{T}{eff}{} + \thee{U}{BO}] \ket{\thewf{nuc}{}{0}} = \thei \hbar \dfrac{\partial}{\partial t}  \ket{\thewf{nuc}{}{0}} .
\label{ham_012}
\end{equation}
We expect $\thewf{nuc}{}{0}$ to be sharply peaked around $\braket{\thewf{nuc}{}{0}|\ther{}|\thewf{nuc}{}{0}}$ and all other quantities to be effectively constant around this peak. As a consequence, we can replace $\ther{}$, $\thep{}$, and $\theop{H}{eff}{}$ by their semiclassical counterparts. 

\subsection{Hamilton's Equations of Motion}

From Eq.\,\eqref{ham_012}, we obtain the following classical Hamiltonian
\begin{equation}
\theop{H}{eff}{} (\ther{},\thep{}) = 
\sum_{\indj=1}^{\then{nuc}} \dfrac{1}{2 \them{\indj}} [\thep{\indj} - \thew{\indj}{\ther{}}]^2 + \thee{U}{BO} (\ther{}) ,
\label{lag_000}
\end{equation}
with the effective total vector potential
\begin{equation}
\thew{\indi}{\ther{}} = \theza{\indi}{\ther{}} - \thechi{\indi}{\ther{}} .
\label{lag_001}
\end{equation}
A disadvantage of the resulting Hamilton's equations of motion,
\begin{align}
\therdot{\indi} &= \dfrac{\partial \theop{H}{eff}{} (\ther{},\thep{})}{\partial \thep{\indi}} = 
\dfrac{1}{\them{\indi}} [\thep{\indi} - \thew{\indi}{\ther{}}] ,\label{hem_000}\\
\thepdot{\indi} &= -\dfrac{\partial \theop{H}{eff}{} (\ther{},\thep{})}{\partial \ther{\indi}}  ,
\label{hem_001}
\end{align}
is that one has to propagate the gauge dependent canonical momenta $\thep{}$. For completeness and to obtain an alternative form of the equations of motion in terms of gauge invariant quantities, we reformulate Eqs.~\eqref{hem_000} and \eqref{hem_001} in the terms of the Lagrangian in the next section.

\subsection{Lagrangian Equations of Motion}

The Lagrangian is obtained by a Legendre transformation of the Hamiltonian in Eq.\,\eqref{lag_000}:
\begin{equation}
\thelag (\ther{},\therdot{}) = 
\max_{\thep{}} \left(
\sum_{\indj=1}^{\then{nuc}} \therdot{\indj} \cdot \thep{\indj} - \theop{H}{eff}{} (\ther{},\thep{}) 
\right).
\label{lag_002}
\end{equation}
In accordance with Eq.~\eqref{hem_000}, the stationary condition on $\thep{}$ gives
\begin{equation}
\thep{} = \them{} \therdot{} + \thew{}{\ther{}} ,
\label{lag_003}
\end{equation}
where $\them{}$ is a diagonal matrix containing all masses $\them{\indi}$.
Upon substitution into the objective function of Eq.\,\eqref{lag_002}, we obtain an explicit expression for the Lagrangian,
\begin{align}
\thelag (\ther{},\therdot{}) = 
\sum \limits_{\indj=1}^{\then{nuc}} \therdot{\indj} \cdot [\dfrac{1}{2} \them{\indj} \therdot{\indj} 
+ \thew{\indj}{\ther{}}] - \thee{U}{BO}(\ther{}) .
\label{lag_004}
\end{align}
By introducing a velocity-dependent potential
\begin{equation}
\thee{\tilde{U}}{} (\ther{},\therdot{}) = 
\thee{U}{BO} (\ther{}) - \sum \limits_{\indj=1}^{\then{nuc}} \therdot{\indj} \cdot \thew{\indj}{\ther{}} ,
\label{lag_005}
\end{equation}
we arrive at our final expression for the Lagrangian
\begin{align}
\thelag (\ther{},\therdot{}) = 
\dfrac{1}{2}\sum \limits_{\indj=1}^{\then{nuc}} \them{\indj} \therdotsq{\indj} - 
\thee{\tilde{U}}{} (\ther{},\therdot{}) .
\label{lag_006}
\end{align}
Before solving the Lagrange equations
\begin{align}
\dfrac{\mathrm d}{\mathrm dt} \dfrac{\partial \thelag (\ther{},\therdot{})}{\partial \therdot{\indi}} &= 
\dfrac{\partial \thelag (\ther{},\therdot{})}{\partial \ther{\indi}} ,
\label{lag_007}
\end{align}
with the Lagrangian of Eq.\,\eqref{lag_006}, we first examine the velocity-dependent potential $\thee{\tilde{U}}{}$ in the next section.

\subsection{Velocity-Dependent Potential}

A central quantity in standard Born--Oppenheimer molecular dynamics is the vector of forces $\thef{}{}{}$, which in the absence of an external magnetic field is obtained as the gradient of $-\thee{U}{BO}(\ther{})$ with respect to $\ther{}$. In our case, this gradient additionally includes a contribution from the effective vector potential:
\begin{align}
-\frac{\partial \thee{\tilde{U}}{} (\ther{},\therdot{})}{\partial \ther{\indi}} &= 
- \frac{\partial \thee{U}{BO} (\ther{})}{\partial \ther{\indi}} 
+ \sum \limits_{\indj=1}^{\then{nuc}}  \bigg[ \therdot{\indj}^\mathrm T \frac{\partial \thew{\indj}{\ther{}}}{\partial \ther{\indi}} \bigg]^\mathrm T \nonumber \\
&= - \frac{\partial \thee{U}{BO} (\ther{})}{\partial \ther{\indi}} 
+ \sum \limits_{\indj=1}^{\then{nuc}} \bigg[\frac{\partial \thew{\indj}{\ther{}}}{\partial \ther{\indi}}\bigg]^\mathrm T \therdot{\indj} \, ,
 \label{vdp_000}
\end{align}
where we have introduced the Jacobian of the effective vector potential $\thew{\indj}{\ther{}}$ with respect to $\ther{\indi}$:
\begin{equation}
\frac{\partial \thew{\indj}{\ther{}}}{\partial \ther{\indi}} =
\begin{pmatrix}
\dfrac{\partial W_{Jx} (\ther{})}{\partial R_{Ix}} & \dfrac{\partial W_{Jx} (\ther{})}{\partial R_{Iy}} & \dfrac{\partial W_{Jx} (\ther{})}{\partial R_{Iz}} \\
\dfrac{\partial W_{Jy} (\ther{})}{\partial R_{Ix}} & \dfrac{\partial W_{Jy} (\ther{})}{\partial R_{Iy}} & \dfrac{\partial W_{Jy} (\ther{})}{\partial R_{Iz}} \\
\dfrac{\partial W_{Jz} (\ther{})}{\partial R_{Ix}} & \dfrac{\partial W_{Jz} (\ther{})}{\partial R_{Iy}} & \dfrac{\partial W_{Jz} (\ther{})}{\partial R_{Iz}}
\end{pmatrix}.
 \label{vdp_001}
\end{equation}
To evaluate the transpose of the Jacobian, we treat the external and internal parts of $\thew{}{\ther{}}$ as given in Eq.\,\eqref{lag_001} separately. Since the nuclear charges and the magnetic field are position independent, the Jacobian of the external vector potential is antisymmetric
\begin{equation}
\frac{\partial \theza{\indj}{\ther{}}}{\partial \ther{\indi}}
=
\dfrac{1}{2}\delta_{\indi\indj} \thez{\indi} e \begin{pmatrix}
0 & -\theb{z}{} & \theb{y}{} \\
\theb{z}{} & 0 & -\theb{x}{} \\
-\theb{y}{} & \theb{x}{} & 0
\end{pmatrix},
 \label{vdp_002}
\end{equation}
so that we obtain the symmetry relation
\begin{align}
\bigg[ \frac{\partial \theza{\indj}{\ther{}}}{\partial \ther{\indi}} \bigg]^\mathrm T  &= -
\frac{\partial \theza{\indi}{\ther{}}}{\partial \ther{\indj}}
 \label{vdp_003} .
\end{align}
A similar symmetry is not found for the geometric vector potential as $\thechi{\indi}{\ther{}}$ depends on the coordinates of all nuclei:
\begin{align}
\bigg[\frac{\partial \thechi{\indj}{\ther{}}}{\partial \ther{\indi}}\bigg]^\mathrm T \neq
-\frac{\partial \thechi{\indi}{\ther{}}}{\partial \ther{\indj}} .
 \label{vdp_004}
\end{align}
Combining Eqs.\,\eqref{vdp_003} and \eqref{vdp_004}, we conclude that
\begin{align}
  \bigg[\frac{\partial \thew{\indj}{\ther{}}}{\partial \ther{\indi}}\bigg]^\mathrm T
= - \frac{\partial \theza{\indi}{\ther{}}}{\partial \ther{\indj}}
-\bigg[\frac{\partial \thechi{\indj}{\ther{}}}{\partial \ther{\indi}}\bigg]^\mathrm T
.
 \label{vdp_005}
\end{align}
Inserting this expression in Eq.\,\eqref{vdp_000}, we obtain the following expression
for the negative gradient of the velocity-dependent potential for nucleus $I$:
%leads to the appearance of the time-derivative of $\theza{\indi}{\ther{}}$:
%
\begin{align}
-\frac{\partial \thee{\tilde{U}}{} (\ther{},\therdot{})}{\partial \ther{\indi}}
&= -\frac{\partial \thee{U}{BO} (\ther{})}{\partial \ther{\indi}} - \thezadot{\indi}{\ther{}}
\nonumber \\
&\quad- \sum \limits_{\indj=1}^{\then{nuc}} \bigg[\frac{\partial \thechi{\indj}{\ther{}}}{\partial \ther{\indi}}\bigg]^\mathrm T \therdot{\indj}
\label{vdp_006}
\end{align}
with a contribution from the time derivative of the magnetic vector potential at the position of nucleus $I$ and a contribution that depends on the velocity of each nucleus in the system.

\subsection{Equations of Motion}

Inserting into Lagrange's equations of motion given in Eq.\,\eqref{lag_007} the Lagrangian in Eq.\,\eqref{lag_006} and using the expression for the negative gradient of the velocity-dependent potential in Eq.\,\eqref{vdp_006}, we obtain
\begin{align}
&\them{\indi} \therddot{\indi} + \thewdot{\indi}{\ther{}} = \nonumber \\
&- \dfrac{\partial \thee{U}{BO}(\ther{})}{\partial \ther{\indi}}
- \thezadot{\indi}{\ther{}} 
- \sum \limits_{\indj=1}^{\then{nuc}} \bigg[\frac{\partial \thechi{\indj}{\ther{}}}{\partial \ther{\indi}}\bigg]^\mathrm T \therdot{\indj} .
\label{eom_000}
\end{align}
Next, introducing the force $\thef{\indi}{}{}=\them{\indi} \therddot{\indi}$ and writing the time-derivative of $\thew{\indi}{\ther{}}$ in terms of its magnetic and geometric components according to Eq.\,\eqref{lag_001}, we find
\begin{align}
\thef{\indi}{}{}
&= - \dfrac{\partial \thee{U}{BO}(\ther{})}{\partial \ther{\indi}} - 2 \thezadot{\indi}{\ther{}} \nonumber \\
&\quad+ \sum \limits_{\indj=1}^{\then{nuc}} 
\frac{\partial \thechi{\indi}{\ther{}}}{\partial \ther{\indj}} \therdot{\indj} 
- \sum \limits_{\indj=1}^{\then{nuc}} 
\bigg[\frac{\partial \thechi{\indj}{\ther{}}}{\partial \ther{\indi}}\bigg]^\mathrm T \therdot{\indj} 
.
\label{eom_001}
\end{align}
Time differentiation of $\thea{\indi}{\ther{}} = \frac{1}{2} \theb{} \times (\ther{\indi}  - \theg{})$ leads to the Lorentz force
\begin{equation}
- 2 \thezadot{\indi}{\ther{}} = Z_I e \therdot{\indi} \times \theb{} ,
\label{eom_002}
\end{equation}
whereas, for the geometric vector potential, we introduce the Berry curvature\cite{Ceresoli2007,Resta2000}
\begin{align}
\theom{\indi}{\indj}{\ther{}} &= 
\dfrac{\partial \thechi{\indi}{\ther{}}}{\partial \ther{\indj}} -
\bigg[\dfrac{\partial \thechi{\indj}{\ther{}}}{\partial \ther{\indi}} \bigg]^\mathrm T , 
\label{eom_003}
\end{align}
whose $\alpha\beta$ Cartesian component is given by\cite{Culpitt2021}
\begin{align}
\Omega^{\indi\indj}_{\alpha\beta} =
\thei \hbar \bigg[
\braket{ \frac{\partial \thewf{el}{}{}}{\partial R_{\indi\alpha}} | 
\frac{\partial \thewf{el}{}{}}{\partial R_{\indj\beta}} } - 
\braket{ \frac{\partial \thewf{el}{}{}}{\partial R_{\indj\beta}} | 
\frac{\partial \thewf{el}{}{}}{\partial R_{\indi\alpha}} }
\bigg],
\label{eom_003b}
\end{align}
and obtain
\begin{align}
\sum \limits_{\indj=1}^{\then{nuc}} 
\bigg \{
\frac{\partial \thechi{\indi}{\ther{}}}{\partial \ther{\indj}}
- 
\bigg[\frac{\partial \thechi{\indj}{\ther{}}}{\partial \ther{\indi}}\bigg]^\mathrm T \bigg \} \therdot{\indj} 
= \sum \limits_{\indj=1}^{\then{nuc}} \theom{\indi}{\indj}{\ther{}} \therdot{\indj} .
\label{eom_004}
\end{align}
Combining Eq.\,\eqref{eom_001} with Eqs.\,\eqref{eom_002} and \eqref{eom_004}, we arrive at the final expression for the force acting on nucleus $\indi$:
\begin{align}
\thef{\indi}{}{}  = 
 - \dfrac{\partial \thee{U}{BO}(\ther{})}{\partial \ther{\indi}} +
\thez{\indi} e \therdot{\indi} \times \theb{} 
+ \sum \limits_{\indj=1}^{\then{nuc}} \theom{\indi}{\indj}{\ther{}}  \therdot{\indj} .
\label{eom_005}
\end{align}
The total force on each nucleus is thus obtained as a sum of three contributions: the Born--Oppenheimer force, the Lorentz force, and the Berry force:
\begin{equation}
\thef{\indi}{}{(\ther{},\therdot{})} = \thef{\indi}{BO}{} (\ther{}) + \thef{\indi}{L}{} (\ther{},\therdot{})  + \thef{\indi}{B}{} (\ther{},\therdot{}) .
\label{eom_006}
\end{equation}
The Berry force represents the screening of the magnetic field by the electrons, which we denote Berry screening from now on. Like the velocity-dependent potential, $\thee{\tilde{U}}{}(\ther{},\therdot{}$), the forces show a velocity dependence, which is not present in the field-free case.

\subsection{Conservation of Energy and Pseudomomentum}

From the physical momenta $\thepi{I}$, which are related to the canonical 
momenta $\thep{\indi}$ in the manner
\begin{align}
\thepi{I} &= \them{I} \therdot{I} = \thep{I} - \thew{I}{\ther{}},
\label{con_000}
\end{align}
we obtain the total energy of the system as
\begin{equation}
\thee{E}{tot} = 
\sum \limits_{\indi=1}^{\then{nuc}} \dfrac{\Pi_{\indi}^2}{2\them{\indi}} + \thee{U}{BO} (\ther{}) .
\label{con_001}
\end{equation}
Please note that our notation for the physical momenta differs from the notation in ref.~\onlinecite{Culpitt2021} ($\thepi{}$ instead of $\bar{\thepi{}}$). While $\thee{E}{tot}$ is conserved (since the Lagrangian has no explicit time dependence), the total canonical and physical momenta are not:
\begin{equation}
\dfrac{\mathrm{d} \thee{E}{tot}}{\mathrm{d}t} = 0,\qquad 
\sum \limits_\indi \dot{\thep{\indi}} \neq 0, \qquad 
\sum \limits_\indi \thepidot{\indi} \neq 0 .
\label{con_002}
\end{equation}
This nonconservation of momenta arises since momenta and coordinates are coupled in the Lagrangian, which is therefore not translationally invariant. Instead, the Lagrangian satisfies the condition
\begin{align}
\sum \limits_{\indi} \frac{\partial \thelag (\ther{},\therdot{})}{\partial \ther{\indi}} &= - \sum \limits_{\indi} \frac{\partial \thee{U}{BO} (\ther{})}{\partial \ther{\indi}} 
- \sum \limits_{\indi} \thezadot{\indi}{\ther{}}
\nonumber \\
&\quad- \sum \limits_{\indi\indj=1}^{\then{nuc}} \bigg[\frac{\partial \thechi{\indj}{\ther{}}}{\partial \ther{\indi}}\bigg]^\mathrm T \therdot{\indj}
\label{con_003}
\end{align}
As the Born--Oppenheimer contribution vanishes due to translational invariance
\begin{align}
\sum \limits_{\indi} \frac{\partial \thee{U}{BO} (\ther{})}{\partial \ther{\indi}} = 0 ,
\label{con_004}
\end{align}
we arrive at the following translational symmetry of the the Lagrangian:
\begin{align}
\sum \limits_{\indi} \frac{\partial \thelag (\ther{},\therdot{})}{\partial \ther{\indi}} =  - \sum \limits_{\indi} \thezadot{\indi}{\ther{}}
- \sum \limits_{\indi,\indj=1}^{\then{nuc}} \bigg[\frac{\partial \thechi{\indj}{\ther{}}}{\partial \ther{\indi}}\bigg]^\mathrm T \therdot{\indj} .
\label{con_005}
\end{align}
Noting that the canonical momentum is given by
\begin{equation}
\thep{\indi} = \frac{\partial \thelag (\ther{}, \therdot{})}{\partial \therdot{\indi}} ,
\label{con_006}
\end{equation}
we conclude from Lagrange's equations of motion that the time derivative of the total canonical momentum is as follows:
\begin{align}
\sum \limits_{\indi} \thepdot{\indi} &= 
 - \sum \limits_{\indi} \thezadot{\indi}{\ther{}}
- \sum \limits_{\indi,\indj=1}^{\then{nuc}} \bigg[\frac{\partial \thechi{\indj}{\ther{}}}{\partial \ther{\indi}}\bigg]^\mathrm T \therdot{\indj} 
\label{con_007}
\end{align}
Introducing the pseudomomentum 
\begin{align}
\thek{} (\ther{},\therdot{}) = \thep{} + \thew{}{\ther{}},
\label{con_008}
\end{align}
we find that the total pseudomomentum
\begin{align}
0 = 
\dfrac{\mathrm{d}\thek{\mathrm{tot}}}{\mathrm{d}t} &=
\sum \limits_\indi \thekdot{\indi} =
\sum \limits_\indi \thepdot{\indi} + \sum \limits_\indi \thewdot{\indi}{\ther{}} \nonumber \\ &=
- \sum \limits_{\indi,\indj=1}^{\then{nuc}} \bigg\{\bigg[\frac{\partial \thechi{\indj}{\ther{}}}{\partial \ther{\indi}}\bigg]^\mathrm T  + 
 \frac{\partial \thechi{\indi}{\ther{}}}{\partial \ther{\indj}}\bigg\} \therdot{\indj}
\label{con_009}
\end{align}
is conserved when
\begin{align}
\bigg[\frac{\partial \thechi{\indj}{\ther{}}}{\partial \ther{\indi}}\bigg]^\mathrm T  = -
 \frac{\partial \thechi{\indi}{\ther{}}}{\partial \ther{\indj}} ,
\label{con_010}
\end{align}
which is true when investigating atoms or highly symmetric molecules, for example H$_2$. In general, however, this equality does not hold. Due to eq.~\eqref{con_010}, eq.~\eqref{eom_003} reduces to
\begin{align}
\theom{\indi}{\indj}{\ther{}} = 2 \frac{\partial \thechi{\indi}{\ther{}}}{\partial \ther{\indj}} ,
\label{con_011}
\end{align}
and the time derivative of $\thechi{\indi}{\ther{}}$ can be written in terms of $\theom{}{}{\ther{}}$:
\begin{align}
\sum \limits_\indj \theom{\indi}{\indj}{\ther{}} \therdot{\indj} = 
2 \thechidot{\indi}{\ther{}} .
\label{con_012}
\end{align}
Assuming that $\text{d}\theom{}{}{\ther{}}/\text{d}t \approx 0$, we can estimate $\thechi{\indi}{\ther{}}$ as follows:
\begin{align}
2 \thechi{\indi}{\ther{}} \approx
\sum \limits_\indj \theom{\indi}{\indj}{\ther{}} \ther{\indj}
\label{con_013}
\end{align}
Inserting this in eq.~\eqref{con_008}, we can calculate $\thek{}$ as a function of $\ther{}$, $\therdot{}$, and $\theom{}{}{\ther{}}$:
\begin{align}
\thek{\indi} (\ther{},\therdot{}) 
&= \them{\indi} \therdot{\indi} - \thez{\indi} \ther{\indi} \times \theb{} - 
2 \thechi{\indi}{\ther{}} \nonumber \\
&\approx \them{\indi} \therdot{\indi} - \thez{\indi} \ther{\indi} \times \theb{} - 
\sum \limits_{\indj=1}^{\then{nuc}} \theom{\indi}{\indj}{\ther{}} \ther{\indj}
\label{con_014}
\end{align}
Note that it is  possible to express the Lagrangian in terms of the pseudomomentum in the following manner:
\begin{align}
\thelag (\ther{},\therdot{}) 
&= \dfrac{1}{2}\sum \limits_{\indj=1}^{\then{nuc}} \therdot{\indj} \cdot \thek{\indj} (\ther{},\therdot{}) 
- \thee{U}{BO} (\ther{})
\label{con_015}.
\end{align}
In contrast to Eq.\,\eqref{lag_006}, this notation introduces a coordinate-dependent kinetic energy contribution, while the potential energy is velocity independent.

Both $\thek{\mathrm{tot}}$ (for some systems) and $\thee{E}{tot}$ can thus be used as a measure for the stability or correctness of dynamics in a magnetic field. At this point, we want to emphasize that all important quantities can be determined directly from $\ther{}$, $\thepi{}$, and $\theom{}{}{\ther{}}$. In particular, we do not need to calculate the gauge-dependent quantity $\thechi{}{\ther{}}$.

\subsection{Equations of Motion with Auxiliary Coordinates and Momenta}

To conduct \textit{ab-initio} molecular dynamics simulations in a magnetic field, it is necessary to integrate the equations of motion (Eq.\,\eqref{eom_005}). Because of the velocity dependence of the forces, the standard propagators cannot be used. In a molecular mechanics framework, where the Berry force is neglected, it is possible to separate the motion due to $\thef{}{L}{}$ from that due to $\thef{}{BO}{}$ and expand it in a Taylor series, leading to a modified Verlet scheme.\cite{Spreiter1999} To our knowledge, the only simulation including the Berry force $\thef{}{B}{}$ has been performed using a Runge--Kutta scheme.\cite{Ceresoli2007}

Here, we apply an alternative approach, based on Tao's symplectic propagation scheme for non-separable Hamiltonians.\cite{Tao2016} It introduces auxiliary coordinates and momenta ($\thertwo{}$, $\theptwo{}$) in addition to the regular pair ($\ther{}$, $\thep{}$) to form an augmented Hamiltonian
\begin{align}
\theopbar{H}{eff}{}& (\ther{}, \thep{}, \thertwo{}, \theptwo{}) = \theop{H}{1}{eff} (\ther{}, \theptwo{}) + \theop{H}{2}{eff} (\thertwo{}, \thep{}) \nonumber \\
&\qquad\qquad  + \dfrac{1}{2}\gamma \Vert \ther{} - \thertwo{} \Vert_2^2
+ \dfrac{1}{2}\gamma^\prime \Vert \thep{} - \theptwo{} \Vert_2^2
,
\label{acm_000}
\end{align}
consisting of two copies $\theop{H}{1}{eff}$ and $\theop{H}{2}{eff}$ of $\theop{H}{eff}{}$ depending on $(\ther{},\theptwo{})$ and $(\thertwo{},\thep{})$, respectively, and coupling terms with coupling constants $\gamma$ and $\gamma^\prime$. If $(\ther{}$, $\thep{}) \approx (\thertwo{}$, $\theptwo{})$ during the propagation, $\theop{H}{eff}{} (\ther{}, \thep{})$ will be close to the exact solution of the system. 

We have applied Tao's approach within a Lagrangian framework, in which the canonical momenta do not appear in the equations of motion. Applying the constraints only to the coordinates, we obtain the following augmented Lagrangian
\begin{align}
\thelagbar(\ther{},\therdot{},\thertwo{},\thertwodot{}) &= \thelag (\ther{},\thertwodot{})\nonumber \\ &+ \thelag (\thertwo{},\therdot{}) + \frac{1}{2} \gamma \Vert \ther{}-\thertwo{} \Vert^2
,
\label{acm_001}
\end{align}
and solve the Lagrange equations
\begin{align}
\dfrac{\mathrm{d}}{\mathrm{d}t} \dfrac{\partial \thelagbar}{\partial \therdot{\indi}} &= 
\dfrac{\partial \thelagbar}{\partial \ther{\indi}} \label{acm_002} , \\ 
\dfrac{\mathrm{d}}{\mathrm{d}t} \dfrac{\partial \thelagbar}{\partial \thertwodot{\indi}} &= 
\dfrac{\partial \thelagbar}{\partial \thertwo{\indi}} \label{acm_003} .
\end{align}
Whether the neglect of the constraint of $\thep{}$ has an impact on the formal symplecticity of the integrator is still under investigation.  The resulting equations
\begin{align}
\them{\indi} \therddot{\indi} &+ 
\sum \limits_{\indj=1}^{\then{nuc}} 
\dfrac{\partial \thew{\indi}{\thertwo{}}}{\partial \thertwo{\indj}} \thertwodot{\indj} = - \dfrac{\partial \thee{U}{BO} (\ther{})}{\partial \ther{\indi}} \nonumber \\ 
& - \sum \limits_{\indj=1}^{\then{nuc}} 
\dfrac{\partial \thew{\indi}{\ther{}}}{\partial \ther{\indj}} \thertwodot{\indj}  \label{acm_004} 
- \gamma (\ther{\indi} - \thertwo{\indi}), \\
\them{\indi} \thertwoddot{\indj} &+ 
\sum \limits_{\indj=1}^{\then{nuc}} 
\dfrac{\partial \thew{\indi}{\ther{}}}{\partial \ther{\indj}} \therdot{\indj}= - \dfrac{\partial \thee{U}{BO} (\thertwo{})}{\partial \thertwo{\indi}} \nonumber \\ 
&- \sum \limits_{\indj=1}^{\then{nuc}}  
\dfrac{\partial \thew{\indi}{\thertwo{}}}{\partial \thertwo{\indj}} \therdot{\indj} 
+ \gamma (\ther{\indi} - \thertwo{\indi}) \label{acm_005} ,
\end{align}
can be rewritten to the final equations of motion:
\begin{align}
\them{\indi} \therddot{\indi} &= - \dfrac{\partial \thee{U}{BO} (\ther{})}{\partial \ther{\indi}} + \thez{\indi} \thertwodot{\indi} \times \theb{} \nonumber \\
&+ \sum \limits_{\indj=1}^{\then{nuc}} \theom{\indi}{\indj}{\ther{}} \thertwodot{\indj}
- \gamma (\ther{\indi} - \thertwo{\indi}) \nonumber \\
&+ \sum \limits_{\indj=1}^{\then{nuc}} 
\bigg [ \dfrac{\partial \thechi{\indi}{\ther{}}}{\partial \ther{\indj}} 
- \dfrac{\partial \thechi{\indi}{\thertwo{}}}{\partial \thertwo{\indj}} 
\bigg ]\thertwodot{\indj} \label{acm_006} \\ 
\them{\indi} \thertwoddot{\indj} &= - \dfrac{\partial \thee{U}{BO} (\thertwo{})}{\partial \thertwo{\indi}} + \thez{\indi} \therdot{\indi} \times \theb{} \nonumber \\
&+ \sum \limits_{\indj=1}^{\then{nuc}} \theom{\indi}{\indj}{\thertwo{}} \therdot{\indj}
+ \gamma (\ther{\indi} - \thertwo{\indi}) \nonumber \\
&- \sum \limits_{\indj=1}^{\then{nuc}} 
\bigg [ \dfrac{\partial \thechi{\indi}{\ther{}}}{\partial \ther{\indj}} 
- \dfrac{\partial \thechi{\indi}{\thertwo{}}}{\partial \thertwo{\indj}} 
\bigg ] \therdot{\indj} 
 \label{acm_007}
\end{align}
where only the geometric vector potential contributes to the last term in each equation since
\begin{equation}
\dfrac{\partial \theza{\indi}{\ther{}}}{\partial \ther{\indi}} = \dfrac{\partial \theza{\indi}{\thertwo{}}}{\partial \thertwo{\indi}} .
\label{acm_008}
\end{equation}
The geometric vector potential does not satisfy the same symmetry but 
\begin{equation}
 \dfrac{\partial \thechi{\indi}{\ther{}}}{\partial \ther{\indj}} \approx \dfrac{\partial \thechi{\indi}{\thertwo{}}}{\partial \thertwo{\indj}} ,
 \label{acm_009}
\end{equation}
when $\ther{} \approx \thertwo{}$ during the integration. Under this assumption, the final terms of eqs.\,\eqref{acm_006} and \eqref{acm_007} can be neglected, yielding:
\begin{align}
\them{} \therddot{}    &= \thef{}{}{} (\ther{},\thepitwo{}) - \gamma (\ther{} - \thertwo{}), \label{acm_010} \\
\them{} \thertwoddot{} &= \thef{}{}{} (\thertwo{},\thepi{}) + \gamma (\ther{} - \thertwo{}), \label{acm_011}
\end{align}
where $\thepi{}$ and $\thepitwo{}$ are equal to $\them{}\therdot{}$ and $\them{}\thertwodot{}$, respectively.

\subsection{Integrating the Equations of Motion}

\begin{figure}
\centering
\includegraphics[width=0.33\textwidth]{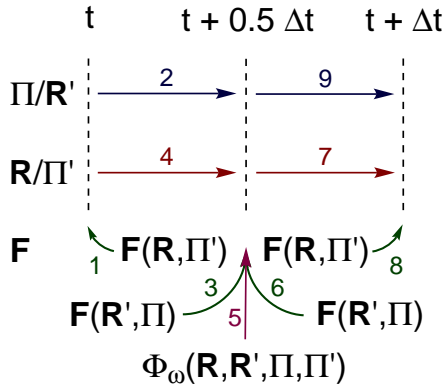}
\caption{Schematic description of the propagation of (auxiliary) nuclear coordinates ($\ther{}$, $\thertwo{}$) and momenta ($\thepi{}$, $\thepitwo{}$) in the auxiliary coordinates and momenta (ACM) method with forces ($\thef{}{}{}$) and the coupling function ($\thephi{\omega}{}$). The numbers illustrate the order of steps in the algorithm.}
\label{auxverlet}
\end{figure}

The integration of eqs.\,\eqref{acm_010} and \eqref{acm_011} can be performed using three types of steps. $\thephi{A}{(\Delta t)}$ involves the propagation of $\thertwo{}$/$\thepi{}$ using $\ther{}$/$\thepitwo{}$
\begin{align}
\thertwo{} (t + \Delta t) &= \thertwo{} (t) + \Delta t \, \them{}^{-1} \thepitwo{} \label{int_000} ,\\
\thepi{}   (t + \Delta t) &= \thepi{}(t) - \Delta t \, \thef{}{}{} (\ther{},\thepitwo{}) \label{int_001} ,
\end{align}
while $\thephi{B}{(\Delta t})$ uses $\thertwo{}$/$\thepi{}$ to propagate $\ther{}$/$\thepitwo{}$
\begin{align}
\ther{} (t + \Delta t) &= \ther{} (t) + \Delta t \, \them{}^{-1} \thepi{}\label{int_002} ,\\
\thepitwo{}(t + \Delta t) &= \thepitwo{}(t) - \Delta t \, \thef{}{}{} (\thertwo{},\thepi{}) \label{int_003} .
\end{align}
The third update ($\thephi{\omega}{(\Delta t)}$) involves the coupling of the two pairs
\begin{equation}
\them{} (\therddot{} - \thertwoddot{}) = - 2 \gamma (\ther{} - \thertwo{}) ,
\label{int_004}
\end{equation}
which results in the following working equations:
\begin{align}
\ther{} (t + \Delta t) 
&= \dfrac{1}{2} \bigg [ [\ther{}(t) + \thertwo{}(t)] + [\ther{}(t) - \thertwo{}(t)] \cos(\theomega{} \Delta t) \nonumber \\
&+ \them{}^{-1}\theomega{}^{-1} [\thepi{}(t) - \thepitwo{}(t)] \sin(\theomega{} \Delta t) \bigg ] \label{int_005} \\
\thertwo{} (t + \Delta t) &= \dfrac{1}{2} \bigg [ [\ther{}(t) + \thertwo{}(t)] - [\ther{}(t) - \thertwo{}(t)] \cos(\theomega{} \Delta t) \nonumber \\
&- \them{}^{-1}\theomega{}^{-1} [\thepi{}(t) - \thepitwo{}(t)] \sin(\theomega{} \Delta t) \bigg ] \label{int_006} \\
\thepi{} (t + \Delta t) &= \dfrac{1}{2} \bigg [ [\thepi{}(t) + \thepitwo{}(t)] + [\thepi{}(t) - \thepitwo{}(t)] \cos(\theomega{} \Delta t) \nonumber \\
&- \them{}\theomega{} [\ther{}(t) - \thertwo{}(t)] \sin(\theomega{} \Delta t) \bigg ] \label{int_007} \\
\thepitwo{} (t + \Delta t) &= \dfrac{1}{2} \bigg [ [\thepi{}(t) + \thepitwo{}(t)] - [\thepi{}(t) - \thepitwo{}(t)] \cos(\theomega{} \Delta t) \nonumber \\
&+ \them{}\theomega{} [\ther{}(t) - \thertwo{}(t)] \sin(\theomega{} \Delta t) \bigg ] \label{int_008}
\end{align}
Here $\theomega{}$ is the coupling strength matrix
\begin{equation}
\theomega{} = \sqrt{2 \gamma \them{}^{-1}},
\label{int_009}
\end{equation}
which is a diagonal matrix with elements $\omega_{\indi}$. As presented in ref.\,\onlinecite{Tao2016}, we can now construct integrators by combining $\thephi{A}{}$, $\thephi{B}{}$, and $\thephi{\omega}{}$. The simplest propagator involves five steps
\begin{align}
\thephi{1}{(\Delta t)} = &\thephi{A}{\bigg(\frac{\Delta t}{2}\bigg)} \circ \thephi{B}{\bigg(\frac{\Delta t}{2}\bigg)} \circ \thephi{\omega}{(\Delta t)} \nonumber \\
&\circ \thephi{B}{\bigg(\frac{\Delta t}{2}\bigg)} \circ \thephi{A}{\bigg(\frac{\Delta t}{2}\bigg)} ,
\label{int_010}
\end{align}
and can be interpreted as an auxiliary coordinates and momenta (ACM) variant of the velocity Verlet\cite{Verlet1967,Swope1982} propagator. The workflow is depicted in Fig.\,\ref{auxverlet}. 

When comparing ACM to the standard velocity Verlet algorithm, there are three differences: (1) $\thertwo{}$ and $\thepitwo{}$ are propagated along with $\thepi{}$ and $\ther{}$, respectively, (2) the propagation of $\ther{}$/$\thepitwo{}$ is divided into two steps, so that $\thephi{\omega}{}$ can be applied when all four components are at the same time step, and (3) three force calculations are required per step. The latter makes the ACM  propagator three times more expensive than the standard propagator.

Higher-order algorithms can be constructed as combinations of the first-order propagator $\thephi{1}{}$ in the manner
\begin{align}
\thephi{n}{(\Delta t)} =& \prod \limits_i^n \thephi{A}{(\Delta t \cdot a_i)} \circ \thephi{B}{(\Delta t \cdot k_i)} \circ \thephi{\omega}{(\Delta t \cdot b_i)} \nonumber \\
&\circ \thephi{B}{(\Delta t \cdot k'_i)} \circ \thephi{A}{(\Delta t \cdot a_i)} .
\label{int_011}
\end{align}
Here, $a_i$ and $b_i$ are pre-optimized coefficients, which are also used to determine $k_i$ and $k'_i$. A pseudo code for an arbitrary order ACM method can be found in the Supporting Information. In this work, we use coefficients (see Table~S1 in the Supporting Information) that have been optimized for molecular dynamics simulations without magnetic field.

\subsection{Calculation of Rovibrational Spectra}

As in the field-free case,  rovibrational spectra of molecules in a magnetic field can be determined as the Fourier transform of the velocity autocorrelation function obtained from nuclear kinetic momenta along a molecular dynamics path:\cite{Futrelle1971,Gaigeot2007,Thomas2013}
\begin{equation}
I(\nu) \propto \sum \limits_{\indi=1}^{N_\text{nuc}} \dfrac{1}{\them{\indi}} \int \mathrm{d}t
\braket{\thepi{\indi} (\tau) \thepi{\indi} (\tau + t)}_{\tau} \mathrm e^{ -2\pi\thei  \nu t  }
\label{vib_000}
\end{equation}
Unlike static calculations based on second-order derivatives, this approach captures anharmonic features of the spectrum and even vibrational overtones are visible.\cite{Gaigeot2007,Thomas2013}

\section{Computational Details}

All simulations in this work are for the lowest singlet states of He or H$_2$ calculated at the HF/cc-pVDZ\cite{Dunning1989} level of theory with the London program package\cite{London}. We consider two magnetic field strengths (0.1 and 1.0\,$\theb{0}$), using field-free calculations as a reference for H$_2$. The Berry curvature needed for Berry screening is determined from finite differences calculations with a step size of $5 \times 10^{-4}\,a_0$  as presented in ref.\,\onlinecite{Culpitt2021}.

For He simulations, we calculated the energy and Berry curvature once and used these values throughout the dynamics simulations. To reduce the computation time of the H$_2$ simulations, we performed a two dimensional scan along its two internal coordinates, the H--H distance ($d$) and the polar angle towards the magnetic field ($\theta$), storing all necessary quantities for the dynamics (energies, forces, Berry curvature) on disk. The 101 $\times$ 101 points with $\Delta d = 0.006\,a_0$ and $\Delta \theta = \pi/100$ were then used to generate a bivariate spline fit of degree three in both directions. Use of these fits instead of calculating the \textit{ab-initio} forces on the fly significantly reduces the computation time, while introducing an error in the standard derivation of the total energy $\sigma(\thee{E}{tot})$ of Eq.\,\eqref{con_001} below the convergence threshold of $10^{-7}\,\thee{E}{h}$ during the dynamics. 

The initial kinetic energy of He was set to $1000$\,K (we express energies in temperature units) and the atom was propagated for $t_\mathrm{tot} = 20$\,ps  with and without Berry screening. The dynamics simulations of H$_2$ started from the global minimum (see potential energy surfaces in Figs.\,S1+S2 in the Supporting Information) with the molecule oriented parallel to the magnetic field ($\theta = 0$). Initial momenta were first chosen randomly; following the removal of the center-of-mass translational component, the momenta were rescaled to yield an initial kinetic energy of $1000$\,K. By doing the latter, we calculate a set of probable trajectories at the given temperature, but not a Boltzmann average. All simulations were conducted for $t_\mathrm{tot} = 20$\,ps. As indicators of the stability of the integration, we use the change of $\thee{E}{tot}$ as well as the mean averaged error of ($\ther{} - \thertwo{}$) and  ($\thepi{} - \thepitwo{}$) obtained from three independent trajectories with different initial momenta. Since the change of the total pseudomomentum ($\thek{\mathrm{tot}}$) was negligible ($< 10^{-14}$) in all our examples, we do not discuss it in this work. When possible, every trajectory was calculated once with and once without the Berry screening for comparison.

The validation of the propagators was performed using the dynamics of H$_2$. To test the influence of the frequency parameter ($\omega = \omega_{\indi}$) and for comparison with the standard velocity Verlet\cite{Verlet1967,Swope1982} and the Taylor expansion ansatz of ref.\,\onlinecite{Spreiter1999}, the ACM version of the Verlet propagator was used with a step size ($\Delta t$) of 0.02\,fs. For testing and comparison of propagators, different $\Delta t$'s and $\omega$'s ($10^{-7}$, $10^{-3}$, and $10^{-1}$) were combined with ACM propagators of different orders ($n$): VV ($n=1$, velocity Verlet\cite{Verlet1967,Swope1982}), FR ($n=3$, propagator of Forest and Ruth\cite{Forest1990}), OY ($n=4$, propagator of Omelyan and coworkers\cite{Omelyan2003}), RK4 ($n=6$, S$_6$/O4 in ref.\,\onlinecite{Blanes2002}), RK6 ($n=10$, S$_{10}$/O6 in ref.\,\onlinecite{Blanes2002}), and RKN6 ($n=14$, SRKN$_{14}^a$/O14-6 in ref.\,\onlinecite{Blanes2002}). The corresponding coefficients ($a$ and $b$) are given Table~S1 in the Supporting Information. 

For the generation of the trajectories and rovibrational spectra of He and H$_2$, the ACM version of the RK4 propagator ($\omega = 10^{-3}$)  was used with $\Delta t$ = 1.0\,fs for He, $\Delta t$ = 1.0\,fs for H$_2$ in the absence of a magnetic field, $\Delta t$ = 0.9\,fs for H$_2$ with $|\theb{}| = 0.1$\,$\theb{0}$, and $\Delta t$ = 0.6\,fs for H$_2$ with $|\theb{}| = 1.0$\,$\theb{0}$. 

\section{Validation and Performance of ACM Integrators}

\begin{figure*}[h]
\begin{tabular}{ll}
(a) & (b) \\
\includegraphics[width=0.48\textwidth]{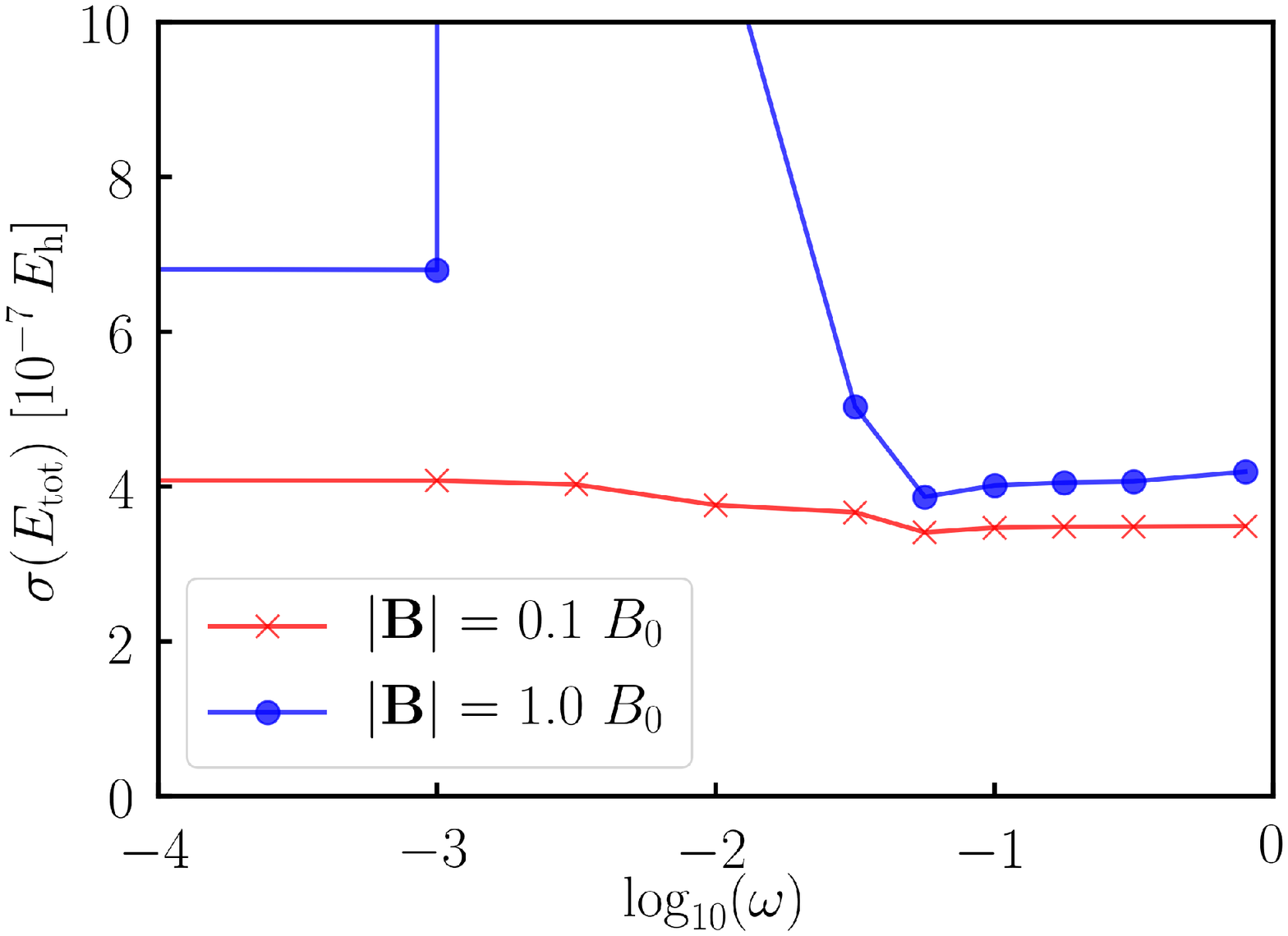} &
\includegraphics[width=0.48\textwidth]{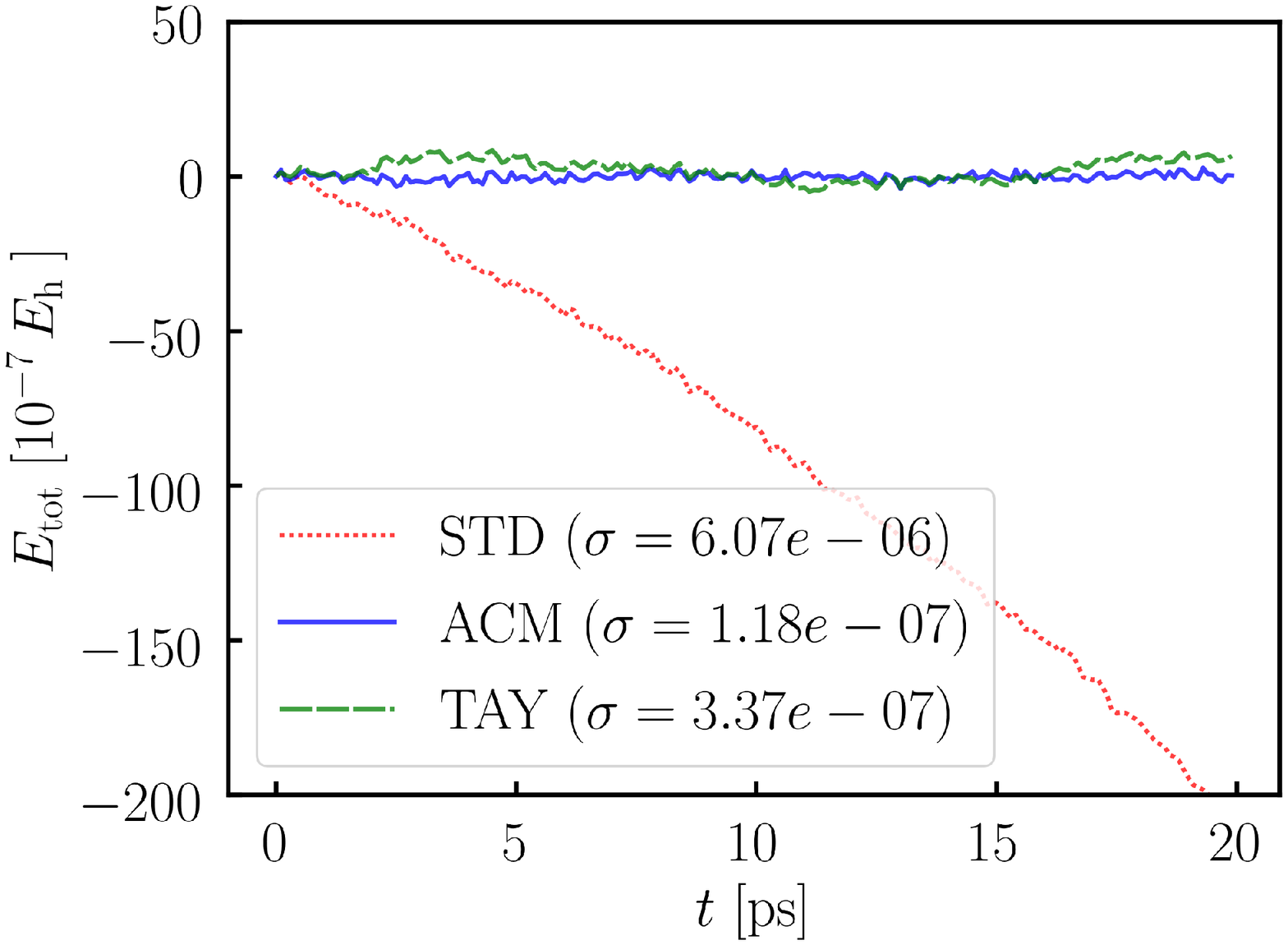} \\
\end{tabular}
\caption{(a) Influence of $\omega$ on the stability of the dynamics of H$_2$ for $|\mathbf{B}|$ = 0.1\,$B_0$ and $|\mathbf{B}|$ = 1.0\,$B_0$ measured by the standard deviation of the total energy ($\sigma(\thee{E}{tot})$). (b) Comparison of integration schemes for one trajectory of H$_2$ with $|\mathbf{B}|$ = 0.1\,$B_0$ using the velocity-Verlet propagator in its standard (STD) implementation, in the auxiliary-coordinates-and-momenta (ACM) form ($\omega = 0.1$), and in its Taylor-expansion (TAY) form. All simulations of (a) and (b) were conducted with and without Berry screening, respectively.}
\label{fig_om_avt}
\end{figure*}

As discussed in Section~IIG, the ACM integrator contains a frequency parameters that couple the two pairs of coordinates ($\ther{}$/$\thertwo{}$) and momenta ($\thepi{}$/$\thepitwo{}$) during the propagation. In Fig.\,\ref{fig_om_avt}a, we show the effect of different values of the coupling constant $\omega = \omega_\indi$ on the standard deviation of the total energy ($\sigma(\thee{E}{tot})$) during simulations with Berry screening. The corresponding plot obtained from simulations without Berry screening as well as  plots showing the averaged error of ($\ther{} - \thertwo{}$) and  ($\thepi{} - \thepitwo{}$) can be found in Figs.\,S5--S7 in the Supporting Information. 

At both  magnetic fields strengths, with or without Berry screening, we observe a region of instability $10^{-3} < \omega < 10^{-1}$. In this range, $\omega$ is close to the frequencies of molecular vibrations and rotations, leading to interference and unstable dynamics. Therefore, we recommend to set $\omega$ to either a larger ($10^{-1}$) or a smaller ($10^{-3}$) value. Figure~\ref{fig_om_avt}a also shows that even smaller $\omega$'s ($10^{-7}$ and $10^{-14}$) can be used to obtain stable dynamics although the \enquote{optimal} value of $\omega$ seems to depend on the propagator and the applied step size, as demonstrated in Fig.\,S9 in the Supporting Information. Whereas an $\omega$ value of 0.1 is ideal for the velocity Verlet propagator with every tested $\Delta t$, the RK4 propagator yields on average better results when setting $\omega$ to $10^{-3}$ or $10^{-7}$. 

Using the optimal value of $\omega = 0.1$, we compare the ACM velocity Verlet integrator with its standard\cite{Verlet1967,Swope1982} and Taylor-expanded\cite{Spreiter1999} variants. In Fig.\,\ref{fig_om_avt}b, we show one trajectory of H$_2$ with $|\theb{}| = 0.1$\,$\theb{0}$. Two additional trajectories with $|\theb{}| = 0.1$\,$\theb{0}$ as well as three trajectories with $|\theb{}| = 1.0$\,$\theb{0}$ are given in Figs.\,S3 and S4 in the Supporting Information. In all simulations, we neglect the Berry force as it cannot be included straightforwardly in the Taylor-expansion scheme. 

The standard velocity-Verlet propagator clearly fails to describe dynamics in a strong magnetic field, exhibiting a systematic drift of $\thee{E}{tot}$ due to the fact that the forces depend on the nuclear velocities.\cite{Spreiter1999} Incorporation of the effect of the magnetic field via a Taylor expansion significantly improves the result, since the energy drift vanishes. The results of the ACM integrator are similar but the error seems to be smaller and more systematic, indicating a better long-term stability. Inclusion of the Berry force has no impact on the stability of the ACM trajectories. 

\begin{figure*}[h]
\begin{tabular}{ll}
(a) $|\mathbf{B}|$ = 0.1\,$B_0$ & (b) $|\mathbf{B}|$ = 1.0\,$B_0$\\
\includegraphics[width=0.48\textwidth]{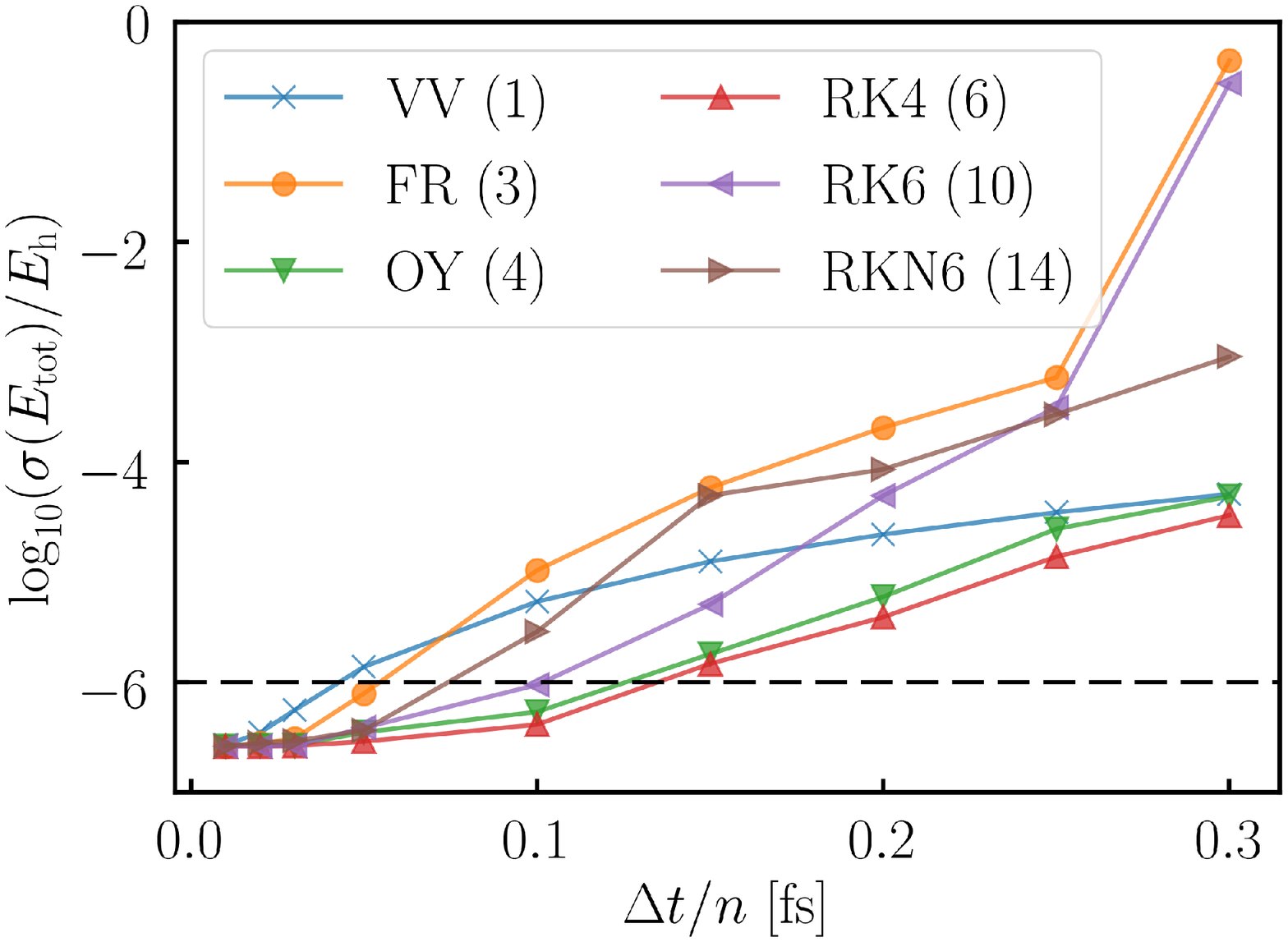} &
\includegraphics[width=0.48\textwidth]{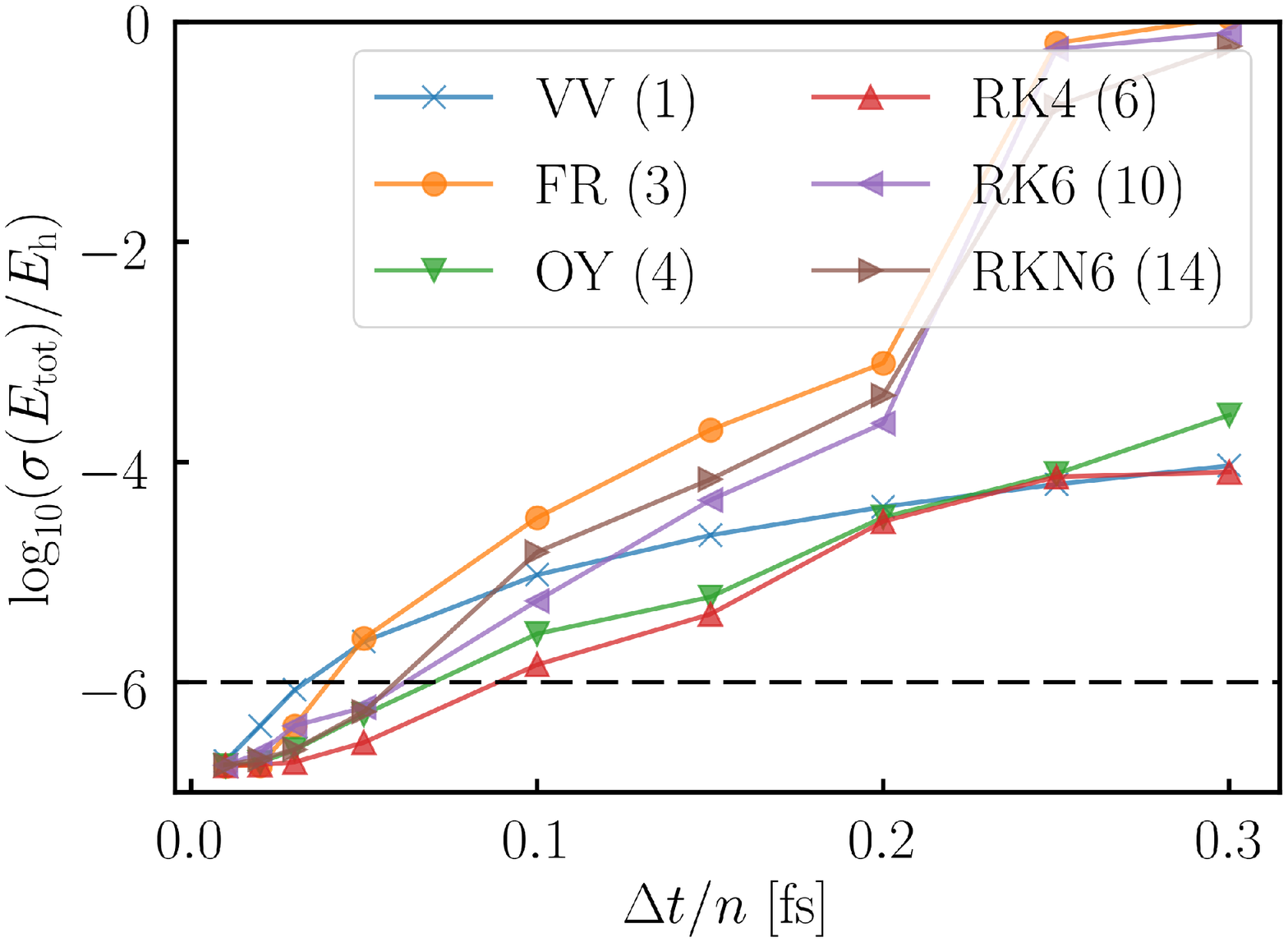} \\
\end{tabular}
\caption{Influence of the propagator and the step size ($\Delta t$) on the stability of the dynamics of H$_2$ for $|\mathbf{B}|$ = 0.1\,$B_0$ (a) and $|\mathbf{B}|$ = 1.0\,$B_0$ (b) with Berry screening, using an \enquote{optimal} value of $\omega$ ($10^{-1}$, $10^{-3}$, or $10^{-7}$). The time step $\Delta t$ is divided by the order ($n$) of the corresponding propagator (see legend) to yield the same computational cost. Our desired standard deviation of the total energy ($\sigma(\thee{E}{tot})$) of $10^{-6}\,\thee{E}{h}$ is indicated by the dashed line.}
\label{fig_prop}
\end{figure*}

Although the ACM integrator appears to yield a better stability than the Taylor-expansion integrator while also enabling the use of the screened Lorentz force, it still requires three times as many force calculations than the other approaches. To improve on this, we have implemented higher-order propagators, which are also used to accelerate field-free simulations.\cite{Forest1990,Omelyan2003,Blanes2002} Figure~\ref{fig_prop} shows the errors of six different propagators with different orders (up to $n = 14$) for different effective step sizes ($\Delta t/n$) using two different magnetic field strengths and an \enquote{optimal} value of $\omega$, obtained by selecting among three values of $\omega$ ($10^{-1}$, $10^{-3}$, and $10^{-7}$) the value that yields the smallest $\sigma(\thee{E}{tot})$ for a given propagator--$\Delta t$ combination. Use of the effective step size $\Delta t/n$ allows for a fair comparison of propagators of different orders, as it accounts for the number of force calculations per step. The results without Berry screening are shown in Fig.\,S8 in the Supporting Information. Details on the propagators are found in Section~III and in Table~S1 in the Supporting Information.

The higher-order ACM propagators significantly improve upon the ACM velocity-Verlet propagator regarding the stability of the dynamics, especially for small $\Delta t/n$ values. The effect seems to be independent of the field strength and inclusion of the Berry screening. In our test set using $|\theb{}|$ = 0.1\,$\theb{0}$ and $|\theb{}|$ = 1.0\,$\theb{0}$, RK4, a partitioned Runge--Kutta propagator with $n = 6$, performs best. Aiming for an accuracy of $\sigma(\thee{E}{tot}) \approx 10^{-6}\,\thee{E}{h}$, it allows for a three times larger $\Delta t/n$ step size than the velocity-Verlet propagator, compensating for the requirement of three forces calculations per step. Consequently, for the simulations of H$_2$ presented in the next section, we have used the RK4 propagator with $\Delta t = 0.9$\,fs in case of $|\theb{}|$ = 0.1\,$\theb{0}$ and $\Delta t = 0.6$\,fs in case of $|\theb{}|$ = 1.0\,$\theb{0}$, with an \enquote{optimal} $\omega$ value of $10^{-3}$.

\section{Illustrative Applications}

\subsection{\enquote{Translational} Spectra of He}

\begin{figure*}[h]
\centering
\begin{tabular}{ll}
(a) & (b) \\
\includegraphics[width=0.48\textwidth]{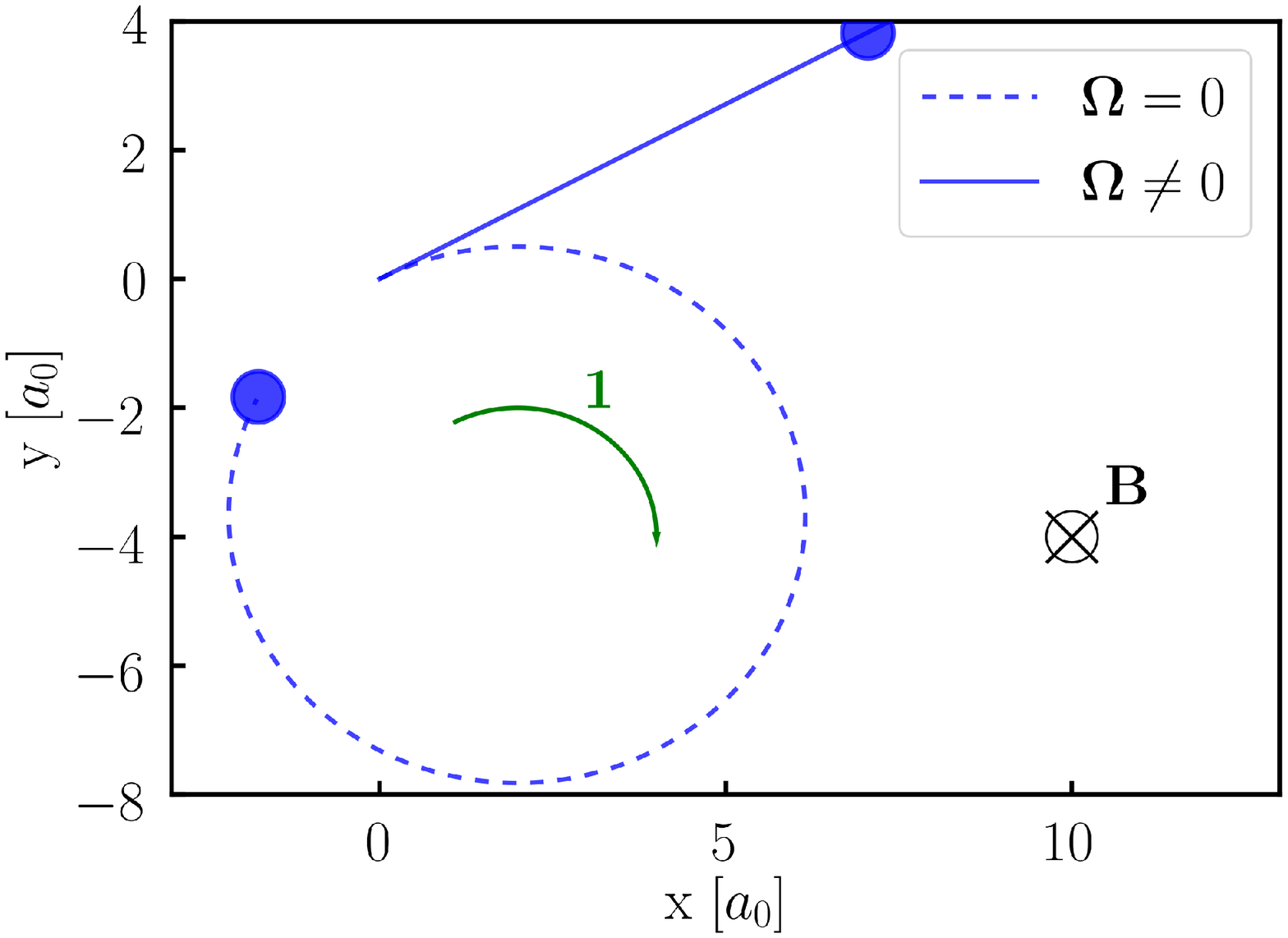} &
\includegraphics[width=0.48\textwidth]{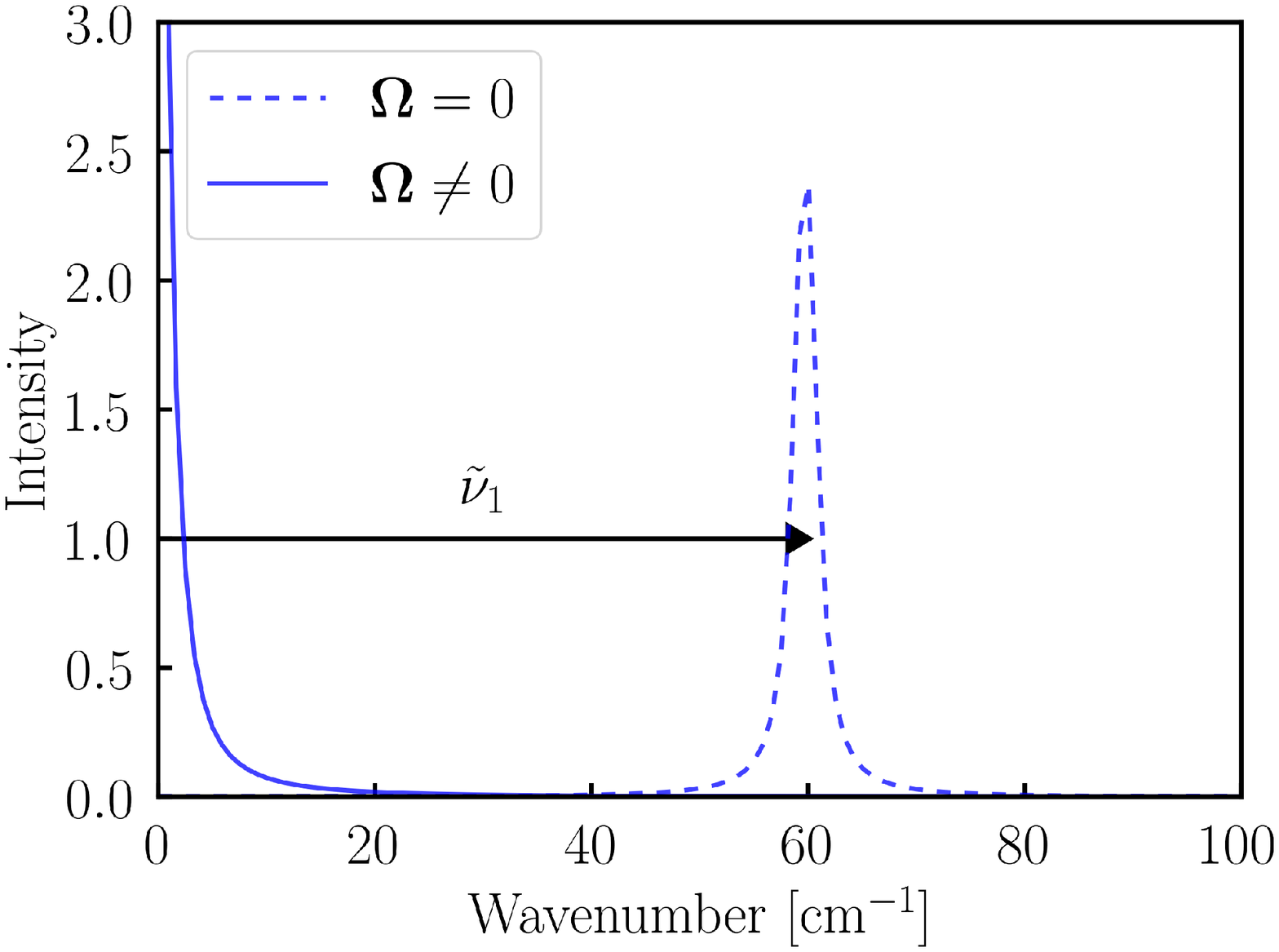} \\
\end{tabular}
\caption{Influence of electron screening of the Lorentz force (Berry force, $\boldsymbol{\Omega}$) on the trajectory (a) and the resulting \enquote{translational} spectrum (b) of He simulated at $|\mathbf{B}|$ = 1.0\,$B_0$. The frequency $\thewvn{1}$ in (b) corresponds to the cyclotron rotation indicated by the arrow in (a).}
\label{fig_he}
\end{figure*}

To illustrate the effect of the screening of the magnetic field by the electrons, we show trajectories and the resulting spectra of He at $|\theb{}|$ = 1.0\,$\theb{0}$ in Fig.\,\ref{fig_he}. The results obtained at $|\theb{}|$ = 0.1\,$\theb{0}$ are found in Fig.\,S10 in the Supporting Information. As indicated by Ceresoli and coworkers\cite{Ceresoli2007}, the Berry force is crucial for obtaining the right physical behavior of an atom in a magnetic field. If it is neglected, the dynamics is dominated by the Lorentz force acting on the bare nuclear charge, leading to a circular motion, which is also visible in the \enquote{translational} spectrum.
%henceforth, we use the index $1$ to describe this cyclotron rotation and its frequency ($\thewvn{1}$). 

For atoms, the Berry force cancels the Lorentz force exactly. Consequently, the initial velocity of He is conserved and we only obtain a peak at 0 cm$^{-1}$ in the translational spectrum, as expected. We note, however that in calculations with atom-fixed Gaussian orbitals, the cancellation of the Lorentz and Berry forces occurs only when London orbitals are used; without London orbitals, the cancellation is incomplete, even in very large basis sets.~\cite{Culpitt2021}

\subsection{Rovibrational Spectra of H$_2$}

As a first molecular application, we conduct \textit{ab-initio} molecular dynamics simulations of H$_2$ in strong magnetic fields ($0.1B_0$ and 1.0\,$\theb{0}$) and use the resulting trajectories to generate rovibrational spectra. The energies of H$_2$ under these conditions depend on the H--H distance ($d$) and the polar angle towards the magnetic field vector ($\theta$). The resulting potential energy surfaces are compared to the surface with $|\theb{}|$ = 0.0 in Figs.\,S1 and S2 in the Supporting Information. The main effects of an increasing magnetic field are that the equilibrium bond distance $d_\text{eq}$ becomes shorter and that the polar rotation (along $\theta$) becomes hindered. Additionally, the equilibrium bond distance $d_\text{eq}$ depends on the polar angle $\theta$. 

The  field strength has a major influence on the trajectories of H$_2$, as seen in Fig.\,S11 in the Supporting Information. While polar rotation is still nearly free at $|\theb{}| = 0.1$\,$\theb{0}$ and an initial kinetic energy of 1000\,K (the barrier height is 0.6 \,m$\thee{E}{h}$), the rotation is not possible at $|\theb{}| = 1.0$\,$\theb{0}$, where the barrier height exceeds 35\,m$\thee{E}{h}$, leading to libration (pendular vibration). In addition, we observe the H--H stretching vibration (along $d$) and the cyclotron rotation discussed in the previous section. In line with the notation introduced there, we use the indices $1$, $2$, and $3$ to describe the cyclotron rotation, the polar rotation/vibration, and the stretching vibration, respectively.

\begin{figure*}[h]
\centering
\includegraphics[width=\textwidth]{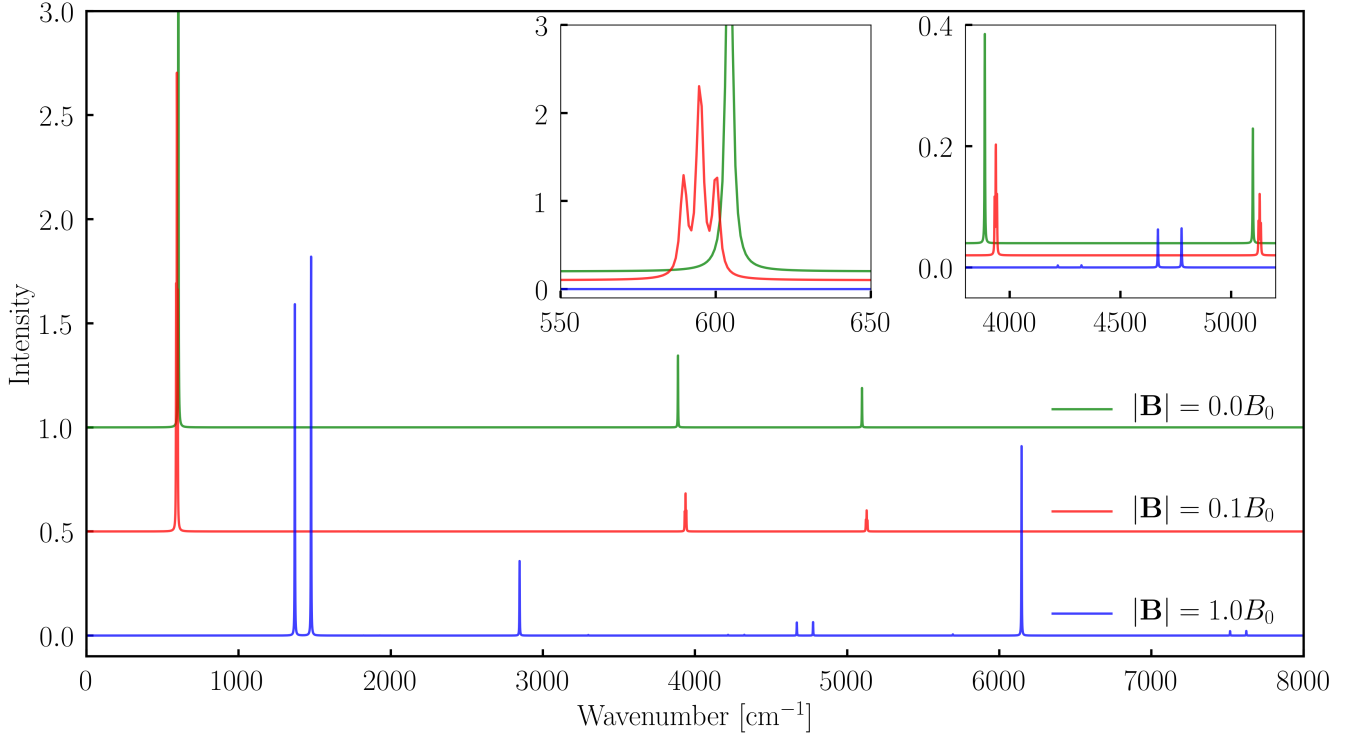} 
\caption{Vibrational spectra obtained from molecular dynamics simulations of H$_2$ with Berry screening at field strengths $|\mathbf{B}|$ = 0.0, 0.1, and 1.0\,$B_0$. The insets show the fine structure of selected regions.}
\label{fig_spec_all}
\end{figure*}

\begin{figure*}[h]
\centering
\begin{tabular}{ll}
(a) $|\mathbf{B}|$ = 0.1\,$B_0$ & (b) $|\mathbf{B}|$ = 0.1\,$B_0$ \\
\includegraphics[width=0.48\textwidth]{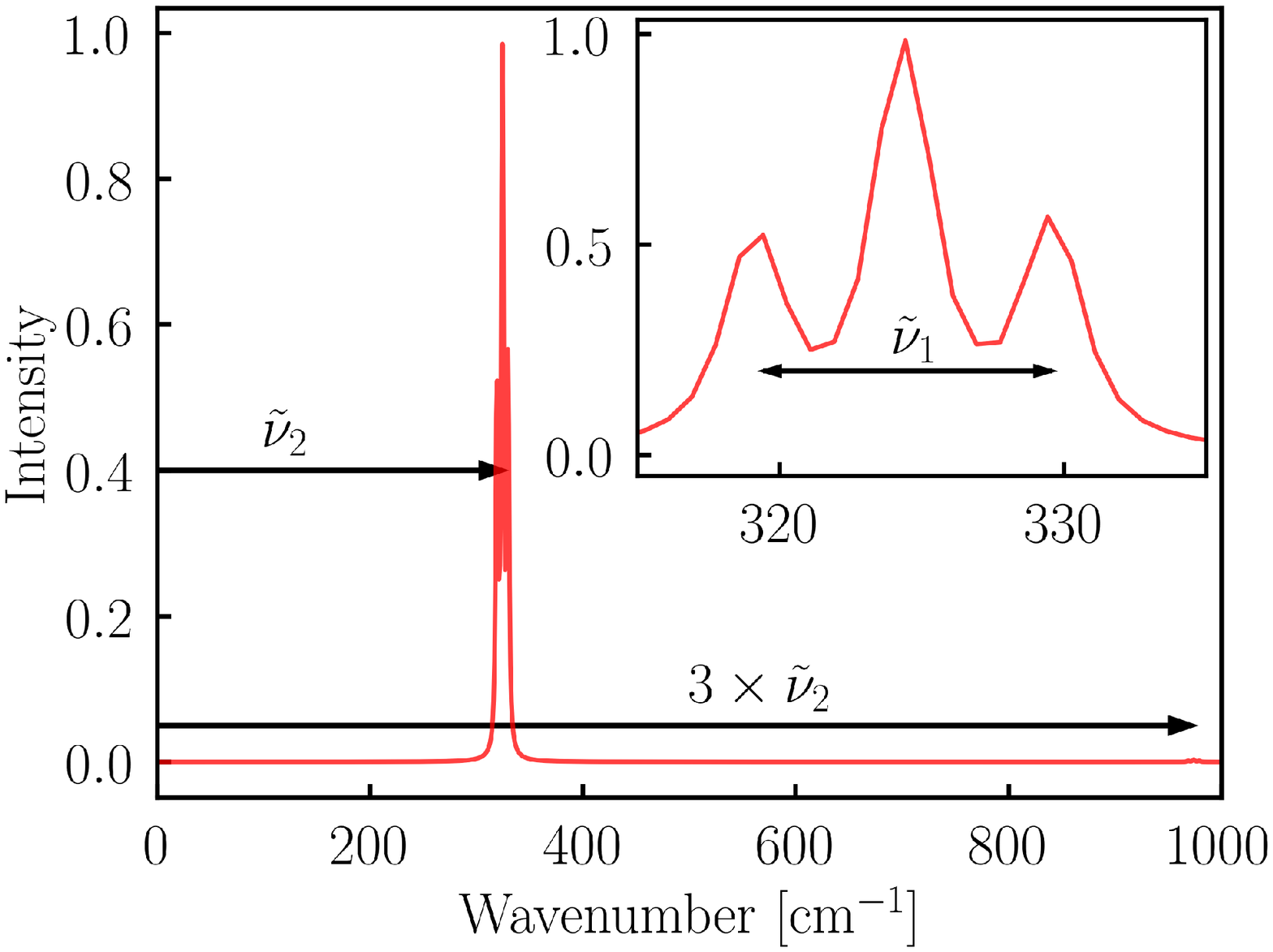} &
\includegraphics[width=0.48\textwidth]{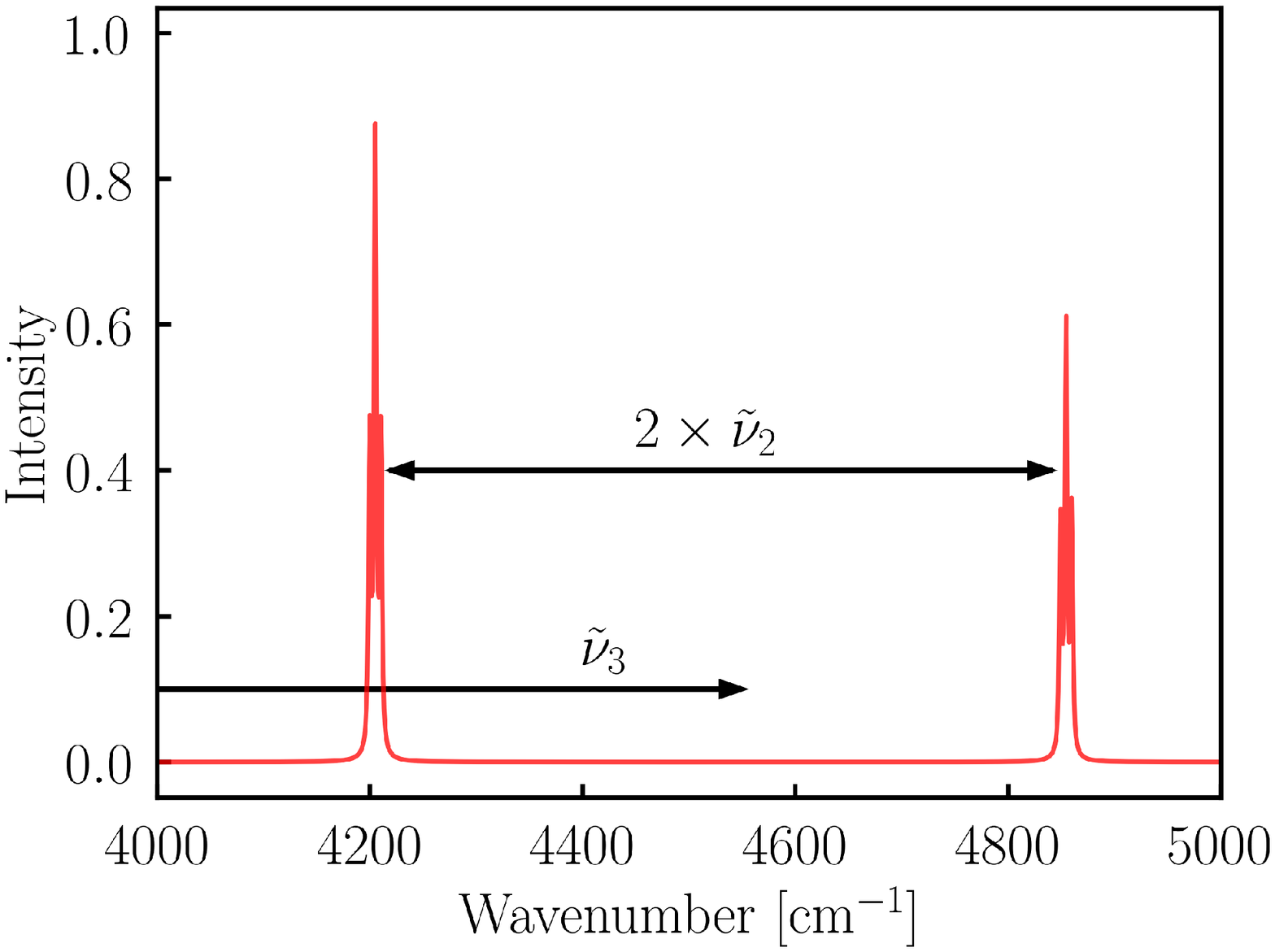} \\
(c) $|\mathbf{B}|$ = 1.0\,$B_0$ & (d) $|\mathbf{B}|$ = 1.0\,$B_0$\\
\includegraphics[width=0.48\textwidth]{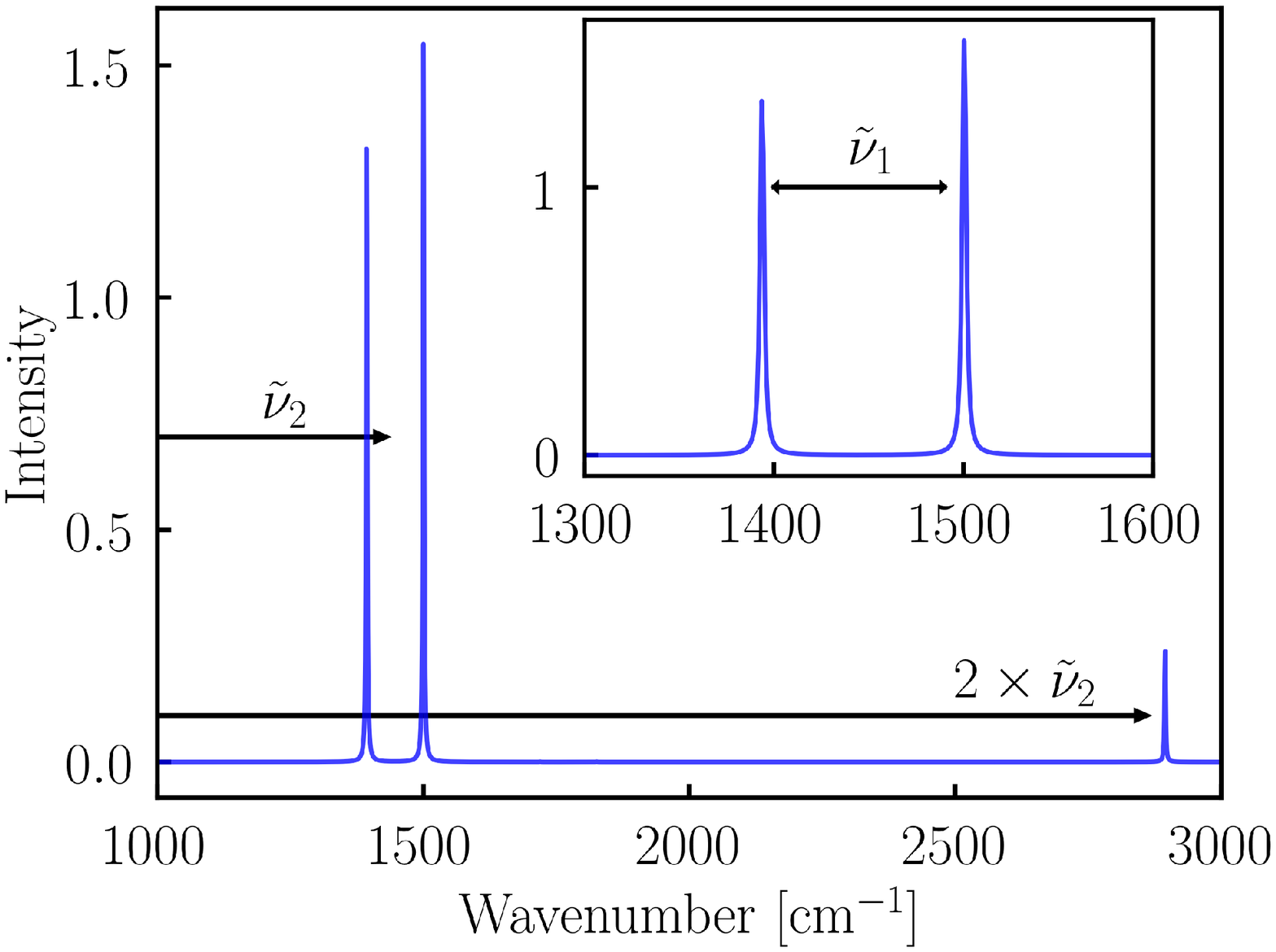} &
\includegraphics[width=0.48\textwidth]{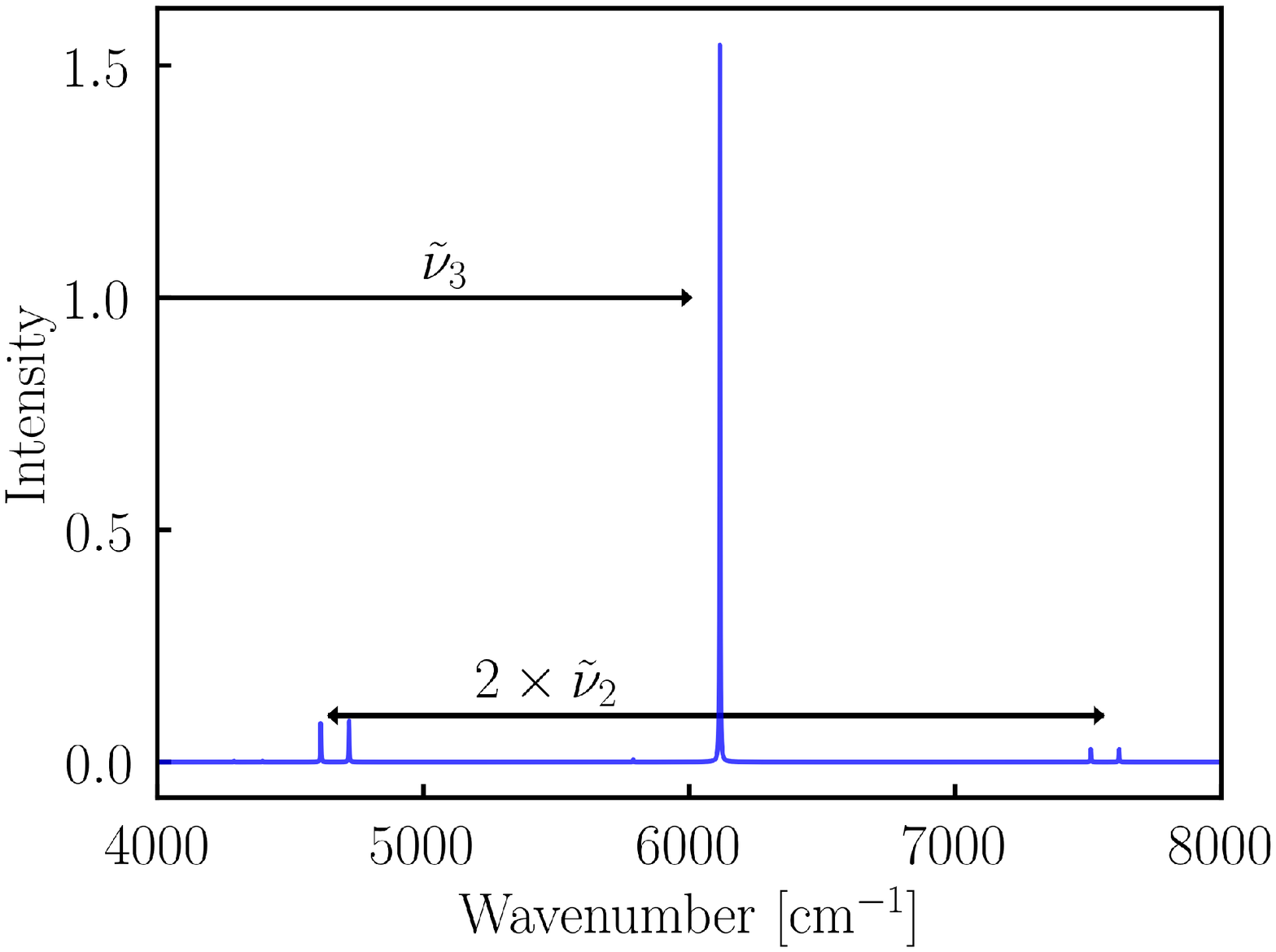} \\
\end{tabular}
\caption{Vibrational spectra obtained from molecular dynamics simulations of H$_2$ with Berry screening at $|\mathbf{B}|$ = 0.1\,$B_0$  (a and b) and  $|\mathbf{B}|$ = 1.0\,$B_0$ (c and d) capturing the lower (a and c) and and higher (b and d) regions of the spectrum. The insets show the fine structure of the first peaks. The frequencies $\thewvn{1}$, $\thewvn{2}$, and $\thewvn{3}$ correspond to the cyclotron rotation, the polar rotation/vibration, and the H--H stretching vibration, respectively.}
\label{fig_spec}
\end{figure*}

In Fig.\,\ref{fig_spec_all}, we compare  rovibrational spectra for the two field strengths to the corresponding field-free spectrum. The fine structure of the spectra as well as the assignment of the vibrations and rotations ($1$, $2$, and $3$) can be seen in Fig.\,\ref{fig_spec}. All H$_2$ spectra obtained with and without Berry screening are also shown in Fig.\,S12 in the Supporting Information. We see immediately that the complexity of the resulting spectra is significantly higher than in the field-free case. Instead of a single peak for the rotation ($\thewvn{2}$) and a doublet for the stretching vibration ($\thewvn{3}$) due to the rotational-vibrational coupling, we observe an entire manifold of peaks and fine structures. 
The peak positions in the field-free spectrum and in the spectrum  at $|\theb{}| = 0.1$\,$\theb{0}$ are very similar---the only differences are that $\thewvn{3}$ is slightly higher (because of stronger binding) and $\thewvn{2}$ slightly lower (because of hindered rotation) in the magnetic field. By contrast, the fine structure is strongly affected by the applied field: as the polar rotation couples to the cyclotron rotation in the magnetic field, $\thewvn{2}$ becomes a triplet with $\thewvn{1}$ as the splitting constant. Additionally, \enquote{rotational overtones} appear at odd multiples of $\thewvn{2}$, featuring the same splitting as the original peak. The fact that \enquote{even rotational overtones} are missing might be a result of the symmetry of the hindrance of the rotation and the resulting selection rules. Note that the position of $\thewvn{2}$ depends solely on the initial velocity (which is the same for all simulations shown in Fig.\,\ref{fig_spec_all}), while its splitting is velocity independent. The peak at $\thewvn{3}$ can be interpreted as a series of doublets created by $\thewvn{2}$ and its fine structure. Each doublet is the result of an individual rotational--vibrational coupling. Consequently, the \enquote{rotational overtones} also couple to $\thewvn{3}$ and vibrational overtones appear at all integral multiplies of $\thewvn{3}$, while showing the same splitting. 

The spectrum at $|\theb{}| = 1.0$\,$\theb{0}$ differs strikingly from the two other spectra. We first note that the polar motion at $\thewvn{2}$ has now become a libration, whose coupling to the cyclotron rotation at $\thewvn{1}$ gives a doublet. Librational overtones appear at every integral multiple of $\thewvn{2}$ but with a twist: even multiples are singlets, while odd multiples preserve the doublet splitting of $\thewvn{2}$. This unusual pattern is again a result of the symmetry and the selection rules of the transitions. The second striking feature of the spectrum at $|\theb{}| = 1.0$\,$\theb{0}$ is the strong blue shift of the stretching vibration $\thewvn{3}$ (reflecting the much stronger binding at this field strength), which now becomes a triplet with $\thewvn{2}$ as the splitting constant. The transition from doublets to triplets occurs as the polar mode $\thewvn{2}$ transmutes from a rotation to a vibration. As in the spectrum at $|\theb{}| = 0.1$\,$\theb{0}$, we observe overtones of $\thewvn{3}$ as well as a coupling of the overtones of $\thewvn{2}$ to $\thewvn{3}$.

\begin{figure*}[h]
\centering
\begin{tabular}{ll}
(a) & (b) \\
\includegraphics[width=0.48\textwidth]{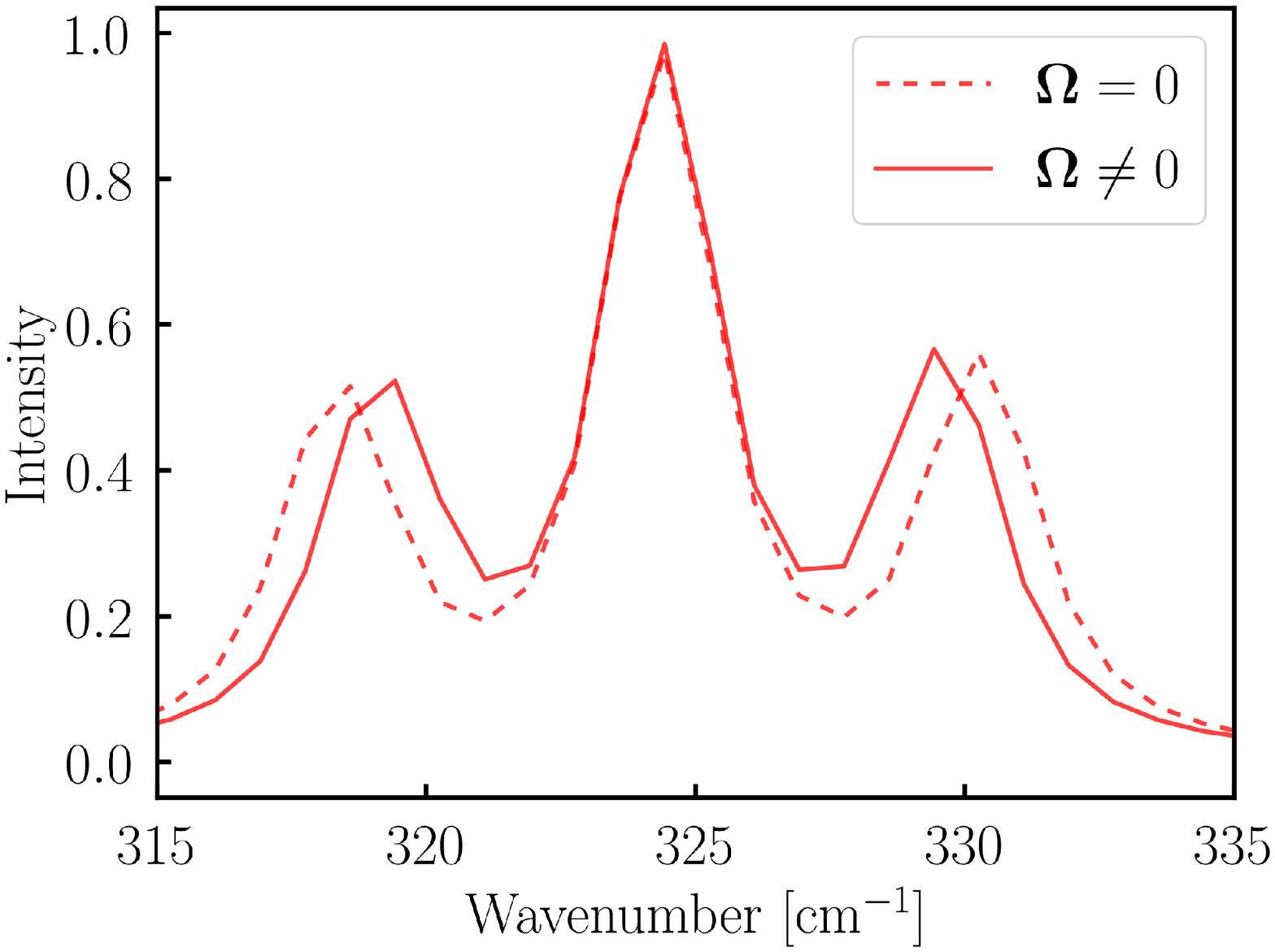} &
\includegraphics[width=0.48\textwidth]{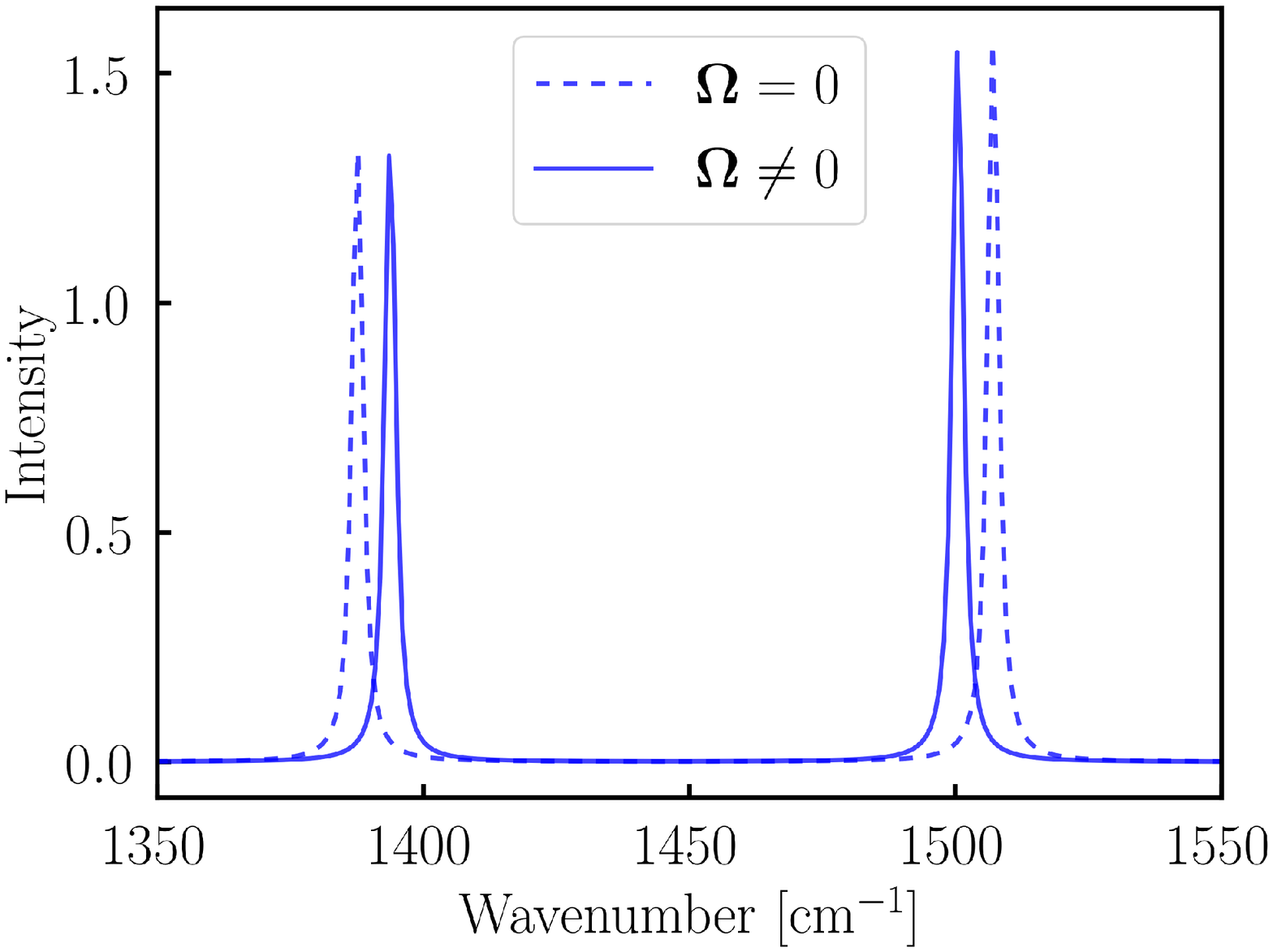} \\
\end{tabular}
\caption{Influence of the Berry screening ($\mathbf{\Omega}$) on the rovibrational spectra of H$_2$ simulated at $|\mathbf{B}|$ = 0.1\,$B_0$ (a) and $|\mathbf{B}|$ = 1.0\,$B_0$ (b).}
\label{fig_screen}
\end{figure*}

When comparing the spectra obtained with and without Berry screening (see Fig.\,S12 in the Supporting Information), the differences are not as severe as in the simulation of He. The main reason for the smaller differences is that we have no center-of-mass motion in our H$_2$ simulations leading to a cyclotron peak in the absence of Berry screening. The only differences between the screened and unscreened spectra are observed for $\thewvn{2}$, which is shown in Fig.\,\ref{fig_screen} for $|\theb{}| = 0.1$\,$\theb{0}$ and $|\theb{}| = 1.0$\,$\theb{0}$. At both field strengths, the position of $\thewvn{2}$ remains unchanged when Berry screening is included, while the splitting is reduced. This behaviour agrees with the observation by Ceresoli et al.\ that the cyclotron rotation is slowed down, as the electrons screen the bare nuclei from the magnetic field.\cite{Ceresoli2007} In our case, this leads to a reduction of the splitting of 10\%. 

\section{Conclusions and Outlook}

The present work contains the first general investigation and application of \textit{ab-initio} molecular dynamics simulations in a strong uniform magnetic field, accounting both for the effect of the field on the electronic structure and for the screening of the Lorentz force by the electrons (Berry screening). Classical trajectories were integrated using a series of newly designed propagators that correctly include the screened Lorentz force in the equations of motion. The precision and performance of the propagators can be gradually tuned by adjusting the step size and/or the order of the propagator. 

As first applications, we simulated the motion of the He atom and the H$_2$ molecule, observed in the atmosphere of nonmagnetic white dwarfs\cite{Xu2013} and speculated to exist also on magnetic white dwarfs. The resulting rovibrational spectra, calculated at field strengths characteristic of magnetic white dwarfs, are surprisingly complex, featuring hindered rotations, librations, as well as unusual splittings and overtones, not present in the field-free case. The Berry screening is essential and cannot be neglected---it is needed to conserve the center-of-mass translation and reduces vibrational level splittings in H$_2$ by up to 10\% compared with corresponding results without screening.

The calculations presented here have revealed many fascinating features of molecular rotations and vibrations in a strong magnetic field but are not quantitative because of the neglect of electron correlation at the Hartree--Fock level of theory and the lack of sampling of initial conditions at a given temperature. Future work comprises a more detailed analysis of selection rules, symmetry, and initial conditions dependence in strong magnetic fields as well as the simulation of larger molecular systems with the inclusion of electron correlation.

\section*{Supplementary Material}

See Supporting Information for potential energy surfaces, implementational details and additional stability analyses of the propagators, and exemplary trajectories as well as full spectra of He and H$_2$.

\section*{Data Availability}

The data that support the findings of this study are available within the article and its Supporting Information.

\section*{Acknowledgements}

We thank Ansgar Pausch and Wim Klopper (Karlsruhe Institue of Technology, KIT) for helpful discussions. This work was supported by the Research Council of Norway through ‘‘Magnetic Chemistry’’ Grant No.\ 287950, and CoE Hylleraas Centre for Molecular Sciences Grant No.\ 262695. This work has also received support from the Norwegian Supercomputing Program (NOTUR) through a grant of computer time (Grant No.\ NN4654K).

\section*{References}

%\bibliographystyle{aipnum4-1}
%\bibliography{aimd_in_b_new,lit}


\end{document}